\documentclass[aps, prb, reprint, superscriptaddress, floatfix, longbibliography]{revtex4-2}

\usepackage{graphicx}
\usepackage{dcolumn}
\usepackage{bm}
\usepackage{amsmath}
\usepackage{amssymb}
\usepackage[caption=false]{subfig}
\usepackage{gensymb}
\usepackage{color}
\usepackage[dvipsnames]{xcolor}
\usepackage{soul} 
\usepackage[english]{babel}

\newcommand{\add}[1]{\textcolor{black}{#1}}
\newcommand{\addB}[1]{\textcolor{black}{#1}}
\newcommand{\addC}[1]{\textcolor{black}{#1}}

\newcommand{\addD}[1]{\textcolor{black}{#1}}

\newcommand{\sub}[1]{}
\newcommand{\subC}[1]{}
\newcommand{\subD}[1]{}

\begin{document}


\title{
The Magnetic Ground State of Atacamite Cu$_2$Cl(OH)$_3$: The Crucial Role of Frustrated Zigzag Chains Revealed by Inelastic Neutron Scattering
}



\author{J. L. Allen}
\email[]{jlallen.correspondence@outlook.com}
\affiliation{Institute for Superconducting and Electronic Materials and School of Physics, University of Wollongong, NSW 2522, Australia}

\author{L. St\"odter}%
\affiliation{Institut für Physik der Kondensierten Materie, Technische Universität Braunschweig, D-38106 Braunschweig, Germany}
\affiliation{Jülich Centre for Neutron Science (JCNS) at Heinz Maier-Leibnitz Zentrum (MLZ), Forschungszentrum Jülich GmbH, 85748 Garching, Germany}

\author{R. A. Mole}
\affiliation{Australian Centre for Neutron Scattering, Australian Nuclear Science and Technology Organisation, \\Lucas Heights, NSW 2234, Australia}

\author{S. S\"ullow} 
\affiliation{Institut für Physik der Kondensierten Materie, Technische Universität Braunschweig, D-38106 Braunschweig, Germany}

\author{O. Janson}
\affiliation{Institute for Theoretical Solid State Physics, Leibniz Institute for Solid State and Materials Research Dresden, 01069 Dresden, Germany}

\author{S. Nishimoto}
\affiliation{Institute for Theoretical Solid State Physics, Leibniz Institute for Solid State and Materials Research Dresden, 01069 Dresden, Germany}
\affiliation{Department of Physics, Technical University Dresden, 01069 Dresden, Germany}

\author{R. A. Lewis}
\affiliation{Institute for Superconducting and Electronic Materials and School of Physics, University of Wollongong, NSW 2522, Australia}

\author{K. C. Rule}
\email[]{kirrilyr@gmail.com\\}
\affiliation{Australian Centre for Neutron Scattering, Australian Nuclear Science and Technology Organisation, \\Lucas Heights, NSW 2234, Australia}
\affiliation{Institute for Superconducting and Electronic Materials and School of Physics, University of Wollongong, NSW 2522, Australia}


\date{\today}

\begin{abstract}
We report inelastic neutron scattering (INS) measurements on the magnetically frustrated $S=\frac12$ sawtooth-chain compound atacamite Cu$_2$Cl(OH)$_3$ featuring inequivalent Cu(1) and Cu(2) sites. Transverse to the sawtooth chains, INS reveals two dispersive spin-wave modes and a gap of at least 0.75\,meV. This behavior is rationalized within a zigzag-chain model of Cu(2) spins in an effective magnetic field of Cu(1) spins. The model is compatible with first-principles calculations and accounts for INS dispersions within linear spin-wave theory calculations. Our results \add{reveal a unique case of an effective separation of energy scales} between \add{two differently oriented one-dimensional chains}, with the zigzag-chain model being essential to fully characterize atacamite's low-energy magnetism.
\end{abstract}


\maketitle

\section{\label{Intro}Introduction}

Frustrated spin systems offer the distinct potential to host novel exotic magnetic phases \cite{Nagler_2017,Balents2010,Bramwell2001,Binder1986} and have become a platform through which to explore fundamental magnetism. The exotic phases emerge from specific magnetic interaction models. For instance, the corner-sharing tetrahedra of pyrochlore lattices can produce spin-ice states \cite{Harris_1997, ramirezZeropointEntropySpin1999, Bramwell2001}, and the kagom\addD{e} lattice is expected to host a quantum spin-liquid state \cite{Balents2010}.
Identifying experimental systems that realize these models allows for rigorous testing of theoretical predictions under real-world conditions.

\addB{Herbertsmithite (ZnCu$_3$(OH)$_6$Cl$_2$) \cite{Norman_2016} is one of the leading candidates for an experimental realization of a quantum spin-liquid material with its geometrically perfect spin-$\frac{1}{2}$ kagom\addD{e} lattice.
It sits within the broader atacamite family of minerals \cite{Malcherek_2018}, with the general formula \subD{(}Zn$_x$Cu$_{4-x}$(OH)$_6$Cl$_2$\subD{)} \cite{Puphal_2018} encompassing polymorphs with various ratios of Zn-Cu substitution.}

\addB{
Atacamite (Cu$_2$Cl(OH)$_3$) refers to the \addC{orthorhombic} polymorph of this mineral family, with full Cu$^{2+}$ occupation.
The spin-$\frac{1}{2}$ Cu$^{2+}$ sites class atacamite as a quantum-magnetic system, and experiments have characterized its geometric frustration effects \cite{Zheng_2005_B}.}
\addB{Furthermore,} atacamite exhibits a complex magnetic phase diagram, including an antiferromagnetic ground state \addB{at $T_\text{N}$ = 8.9\,K \cite{Heinze_2021} (with a magnetic propagation vector of $q$ = [$\frac{1}{2}$, 0, $\frac{1}{2}$] \cite{Heinze_2018} \addC{and with suppressed magnetic mom\subD{e}ents of $0.34\,\mu_\text{B}$ and $0.59\,\mu_\text{B}$ for Cu(1) and Cu(2) sites, respectively})} and \add{a sequence of metamagnetic transitions} up to 30\,T, \subC{including a magnetization plateau \textit{at} 30\,T.} \addC{including a plateau-like \subD{behvaior}\addD{behavior} of the bulk magnetization above 31.5\,T (\textit{H} $\parallel b$ axis) \cite{Heinze_2021}.}
However, \sub{their precise nature}\addB{the precise nature of these transitions and phases} \subC{remains} \addC{remained} ambiguous.

\begin{figure*}[t]
	\centering
	\includegraphics[width=0.9\linewidth,trim={0cm 0cm 0cm 0cm},clip]{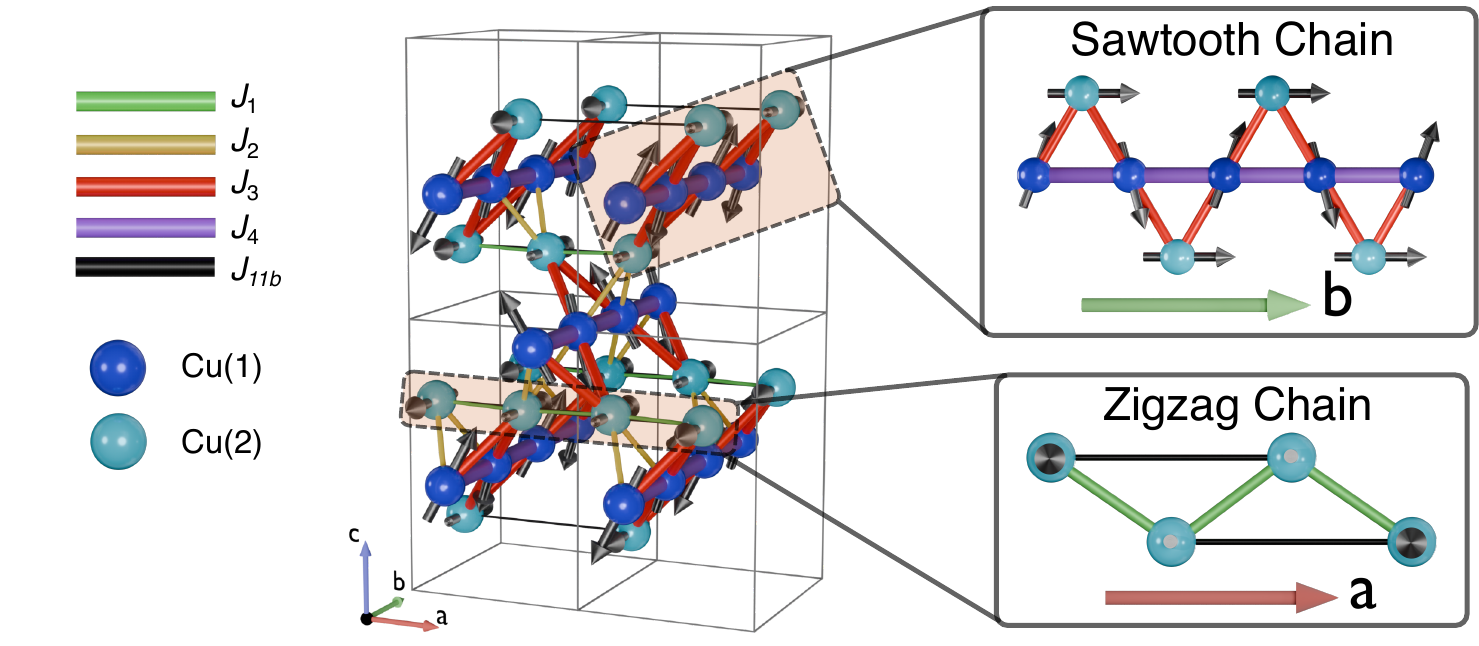}
	\label{Full_dft}
	\caption{ 
		\addB{(Left)} Magnetic\subD{-moment} structure derived from single-crystal neutron diffraction \add{data} \cite{Heinze_2021}, with exchange pathways from the 5 \subD{largest}\addD{strongest} exchange values according to the first-principles analysis in \cite{Heinze_2021}.\subD{of $J_1 = 0.11$\,meV, $J_2 = -0.83$\,meV, $J_3 = 8.79$\,meV, $J_4 = 28.97$\,meV, and $J_{11b} = 1.35$\,meV.
		The structure has refined moment values of $0.34\,\mu_\text{B}$ for the Cu(1) ions, and $0.59\,\mu_\text{B}$ for the Cu(2) ions. However, for LSWT calculations throughout this work normalized moment values have been used, hence ignoring the effect of suppressed moments.}
		\addB{(Right) Expanded in-plane perspectives of the sawtooth-chain and zigzag-chain components in atacamite. The angles between the nearest-neighbor Cu(2) spins in the expanded zigzag chain have been greatly exaggerated for clarity.}
		}
	\label{fig:DFT_Structures}
\end{figure*}

\addB{
Recently, a field-induced quantum critical point at 21.9(1)\,T (\textit{H} $\parallel$ \textit{c} axis) was identified at the \addC{low-temperature} endpoint of the long-range ordered antiferromagnetic phase \cite{heinzeAtacamiteCu22025}. Further, the results \addC{from Ref. \cite{heinzeAtacamiteCu22025}} underpinned the scenario of a dimensional reduction to an effectively one-dimensional behavior occurring at the quantum critical point as the residual 3D exchange couplings are overcome by the magnetic field energy. In this work, we explore the low-energy spin-wave excitations in zero applied field, allowing us to probe in detail the residual exchange couplings in atacamite responsible for the long-range magnetic order below $T_\text{N}$ = 8.9 K and below the critical \addC{magnetic} field.
}

\addB{Regarding the magnetic interactions driving the ground-state magnetism of atacmite,} recent density functional theory (DFT) calculations
\subC{agree well with experimental inverse susceptibility measurements and}
indicate that the dominant exchange interactions form a \add{weak} 3D network of antiferromagnetic $S = \frac12$ sawtooth chains \cite{Heinze_2021}. The five \subD{largest}\addD{strongest} exchange pathways \addD{($J_1 = 0.11$\,meV, $J_2 = -0.83$\,meV, $J_3 = 8.79$\,meV, $J_4 = 28.97$\,meV, and $J_{11b} = 1.35$\,meV)} are shown in Fig. \ref{fig:DFT_Structures}, with the sawtooth chain running along the $b$ direction. Interest has followed atacamite \addB{not only for its \addC{compositional} similarities to Herbertsmithite, but also} due to the unmatched opportunity it offers to \textit{experimentally} investigate the complexities of a low-dimensional quantum spin system within the intensely-studied sawtooth chain model \cite{Hamada_1988,Kubo_1993,Blundell_2003,Hida_2008,Zhitomirsky_2004,Metavitsiadis_2020,Dmitriev_2015,Dmitriev_2018}.

Although the sawtooth chain is the dominant coupling motif in atacamite, DFT calculations also reveal a weaker perpendicular chain \subC{of offset} \addC{formed by competing} nearest and next-nearest interactions \cite{Heinze_2021}. This is the S=$\frac{1}{2}$ zigzag \subD{$J_1-J_2$}\addD{$J_1-J_{11b}$} spin chain \cite{Steven_1996, Aligia_2000, Bursill_1995, Kumar_2015}
(\addD{typically referred to as the $J_1-J_2$ zigzag chain in the literature, and} hereafter referred to as simply the `zigzag chain' to avoid confusion with our exchange-label naming convention),
with Hamiltonian
\begin{equation}
	\mathcal{H}  = \sum_{i=1}^{N}(J_1\mathbf{S}_i\cdot \mathbf{S}_{i+1} + J_{\addD{11b}}\mathbf{S}_i\cdot \mathbf{S}_{i+2}).
\end{equation}
Here, $J_1$ and $J_{\addD{11b}}$ represent the nearest- and next-nearest-neighbor interactions, respectively, on a spin chain with an even number of $N$ spin sites and \add{spin-$\frac{1}{2}$ operators at} spin site $i$.

The zigzag spin chain exhibits diverse magnetic phases dictated by the ratio of $J_{\addD{11b}} / J_1$ \cite{Aguilar_2008,Bursill_1995} owing to the complexity of the competing interactions. These include gapped and gapless spin states, spiral orders, and frustration-driven quantum phases such as spin-nematic states \cite{Bursill_1995, Kumar_2015} and spin-stripe phases. \add{These states have also been revealed for material realizations of the zigzag chain \cite{willenbergComplexFieldInducedStates2016, vekuaCorrelationFunctionsExcitation2007, Pregelj_2015}.}

In this \sub{letter} \addB{investigation}, we demonstrate that atacamite's zigzag chain plays a previously overlooked role in its magnetic behavior at low \add{temperatures}.
We present the first investigation of atacamite's spin-wave excitation spectrum using inelastic neutron scattering (INS) within its ordered phase to directly probe its magnetic interactions. \addD{Similar approaches have been successfully employed to constrain and determine complex Hamiltonians in other $S=\frac12$ zigzag-chain materials \cite{INS_on_zigzag_Pregelj2018, INS_on_zigzag_Hase2024}}.  We find that the zigzag chain is crucial for modeling atacamite, and, when combined with a mean-field treatment, an effective-model Hamiltonian for atacamite is produced that agrees with INS measurements in the low-energy regime. Furthermore, this agreement stands in contrast to the poor match obtained using the full DFT-derived model, whose dominant sawtooth chain is of a \addC{much} higher energy scale than the zigzag chain. An effective separation of energy scales between the two chains is thus presented as \subD{a puzzling}\addD{an interesting} characteristic of a material realization of a frustrated quantum magnet.

The compatibility between a magnetic model and INS measurements can only be justified within a robust theoretical framework. Linear spin-wave theory (LSWT) stands out as a straightforward, common \cite{Zhu_2021}, and reliable method for making quantitative comparisons between a magnetic Hamiltonian and INS experiments in long-range ordered spin systems.

\textcolor{black}{LSWT and INS measurements are often used together to validate first-principles calculations and refine empirical magnetic models, including for frustrated (quantum) spin systems \cite{Harris_1997,Fennell_2004,Haraldsen_2010,Mirebeau_2014,Rule_2017, Heinze_Linarite_2022}.}

The surprising success of LSWT in its application to quantum systems is rooted in the abundance of semiclassical behaviors \cite{richterQuantumMagnetismTwo2004}, even in the $S=\frac12$ limit \cite{ito_structure_2017} \addD{\cite{INS_on_zigzag_Pregelj2018}}. The LSWT calculations within this work were performed using the Sunny package \cite{Sunny} where \add{the magnetic structure refined from neutron diffraction data} \cite{Heinze_2021} was chosen for the spin structure input.

\begin{figure*}[t!]
	\centering

	\subfloat[\label{fig:INS_Data}]{%
 	\includegraphics[width=0.49\textwidth,trim={-1.5cm 0cm -1.5cm 0cm},clip]{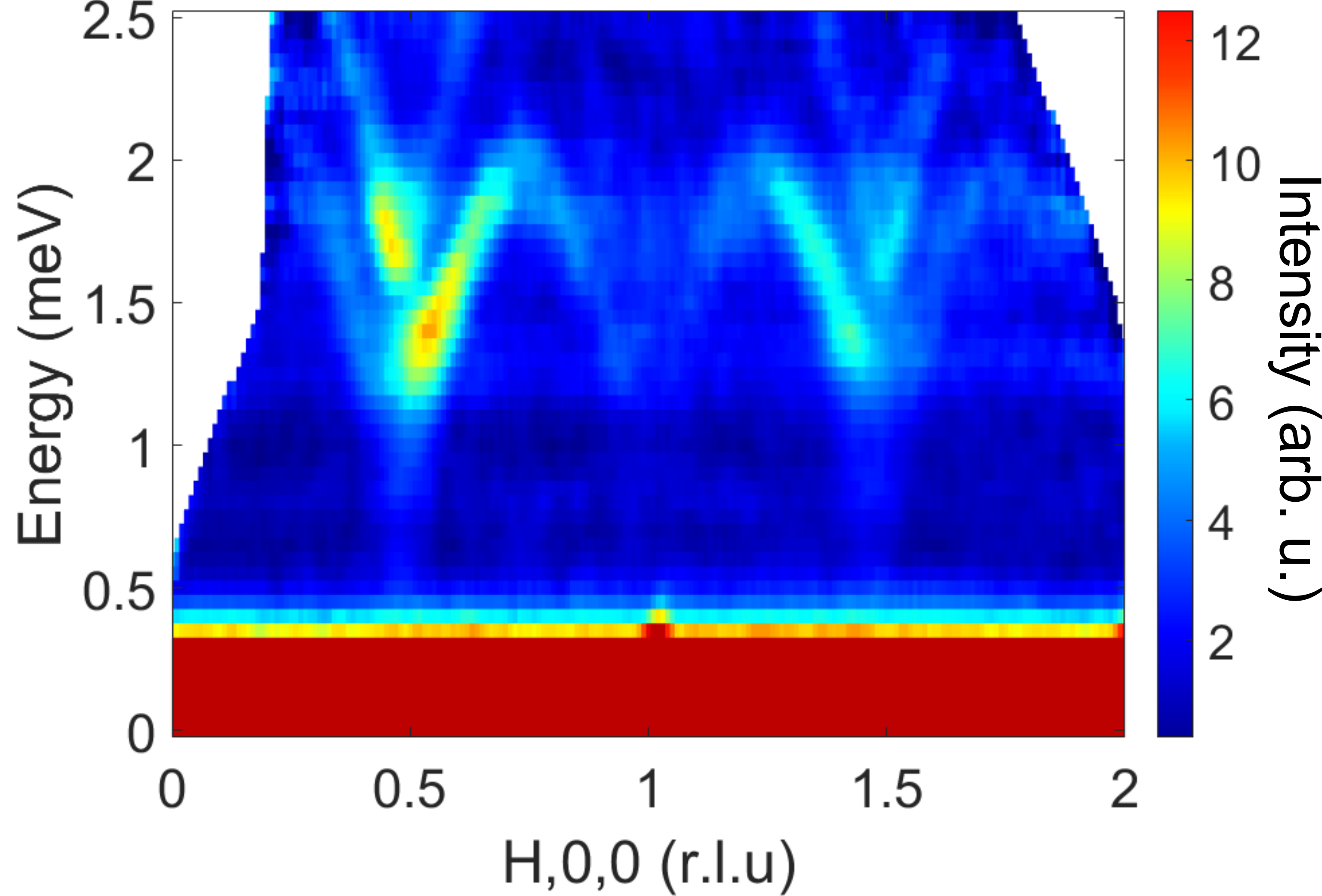}%
		}
	\subfloat[\label{fig:aniso_zigzag}]{%
	\includegraphics[width=0.425\textwidth,trim={15cm 5cm 11cm 2cm},clip]{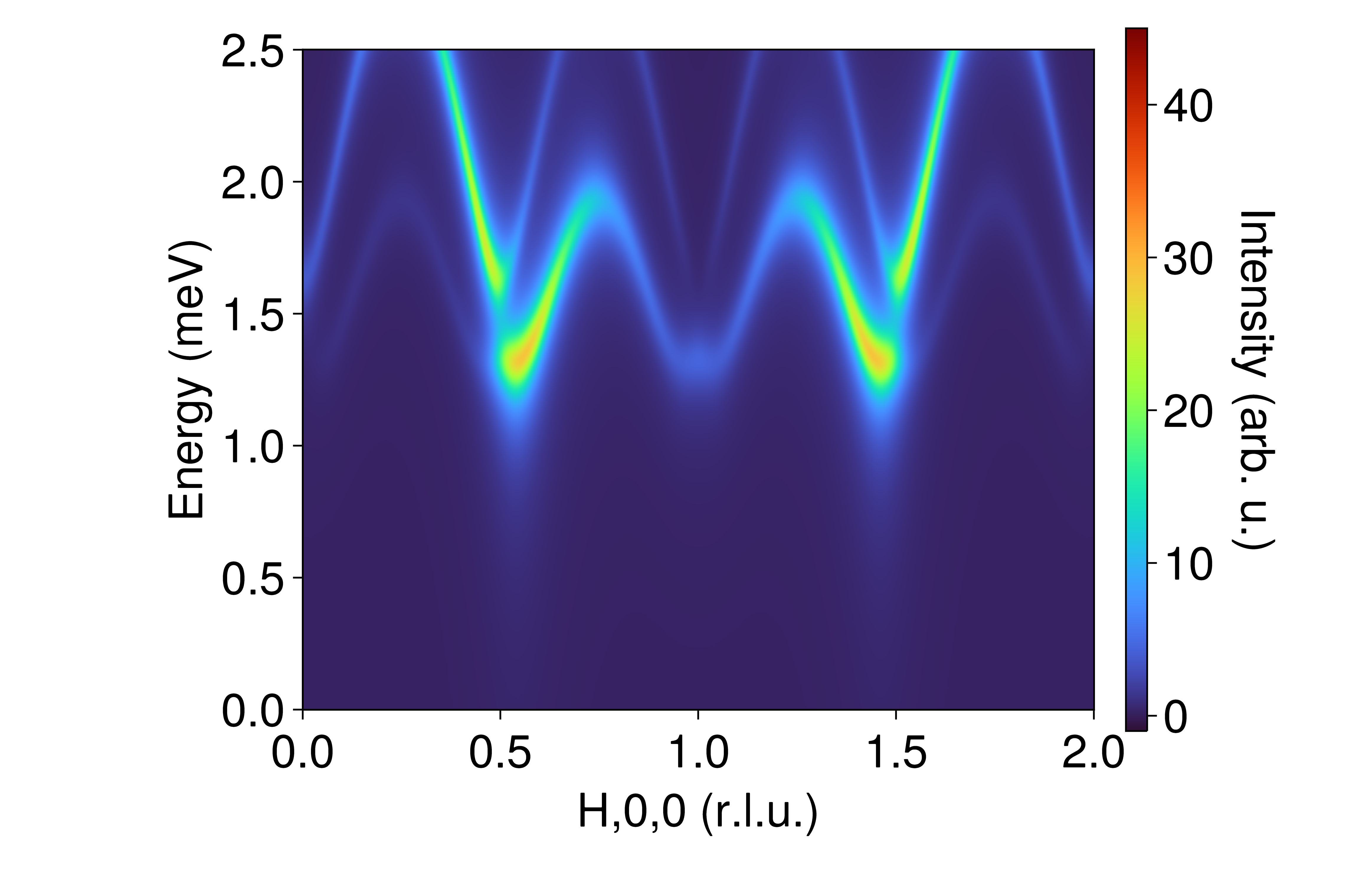}%
		}

	\caption{(a) Experimental INS data of (co-aligned) single-crystalline atacamite measured on Pelican at ANSTO \addC{at 1.5\,K and at zero external field}. Strong dispersion can be observed in the \subD{H}\addD{(H,0,0)} direction with two separate nested modes. (b) LSWT calculations \subC{using Sunny} of the \addD{low-energy} spin-wave modes in atacamite along \subD{the H direction}\addD{(H,0,0)} for a zigzag-chain model \addD{with an effective mean field $h$ = 0.26\,meV as defined in Eq. (\ref{eq:atacamite_hamiltonian})}, \addC{using Sunny \cite{Sunny}}. \addD{Note that normalized moment values have been used in this calculation, hence ignoring the effect of suppressed moments.}}
\end{figure*}


\section{\label{Method}Experimental Methods}

\add{Single-crystal} neutron scattering measurements of atacamite's spin-wave spectrum were performed at the Australian Nuclear Science and Technology Organisation (ANSTO) on the Pelican \cite{PELICAN} time-of-flight INS spectrometer \add{with the sample rotated} in the H0L plane. Pelican was configured with 4.69\,Å neutrons and an approximate resolution of 0.135\,meV at the elastic line. 

The \add{single-crystal} INS data represents a four-dimensional dataset, comprising three spatial dimensions and one energy dimension. Although aligned in the H0L plane, the 0K0 direction could be measured out-of-plane up to $\pm\, 10\degree$ due the large detector height at Pelican of 1\,m, resulting in $\pm\, 0.2$\,r.l.u. of measurable range in K.

Data were collected for the sample at 1.5\,K, 6\,K, and 20\,K in a closed-cycle helium cryostat, and were measured across a sample rotation of $100\degree$ in $1\degree$ intervals. A vanadium sample was used to normalize the detector efficiencies and an empty cryostat scan was used for background subtraction.

Naturally-formed atacamite crystals were measured in this work. Atacamite crystallizes in an orthorhombic \subC{$Pnma$ structure} \addC{crystal structure with space group $Pnma$ and} with $a=6.030(2)$\,\AA, $b=6.865(2)$\,\AA, and $c=9.120(2)$\,\AA \cite{Parise_1986} (see Section I in the supplemental material \cite{supplemental} for more details on the chemical unit cell). \sub{It has a magnetic transition temperature of T$_\text{N}$ = 8.4\,K 
and magnetic propagation vector $q$ = [$\frac{1}{2}$ 0 $\frac{1}{2}$] 
.}

Four single crystals of atacamite with a total mass of 1.8\,g were co-aligned in the H0L plane using the high-intensity diffractometer Wombat \cite{WOMBAT}. Aluminium mounting methods were employed to avoid the effects of hydrogen \add{from glue} \cite{Rule_Glue}.

\section{\label{Results}Results}

\begin{figure}[b!]
	\centering
	\includegraphics[width=0.49\textwidth,trim={0cm 0cm 0cm 0cm},clip]{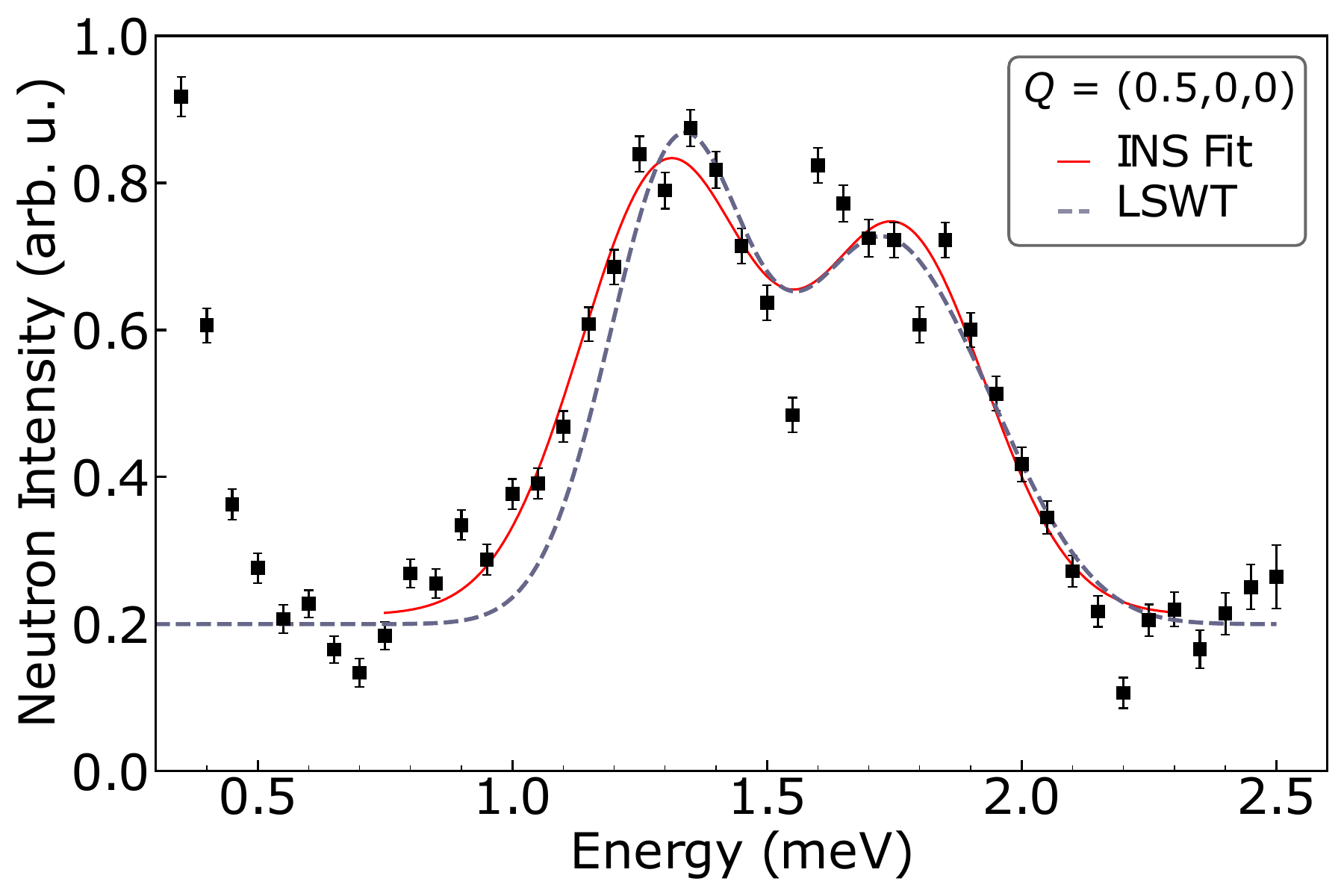}
	\caption{\subD{A 1D cut of the INS data along the H direction at Q = 0.5\,r.l.u..}
			\addD{A constant-$Q$ INS spectrum at (0.5,0,0).} 
			An integration width of Q = 0.04\,r.l.u. has been used for the experimental data. \subD{whereas Q = 0.15\,r.l.u. is the LSWT width in H}\addD{A width of Q = 0.15\,r.l.u. has been used for the LSWT-calculated data} to mimic experimental broadening.
			\addD{The} red line is a double Gaussian fit to the INS data and the \addD{dashed} blue line is the Sunny LSWT calculation, equivalent to that in Fig. \ref{fig:aniso_zigzag}.
			For the lower-energy INS mode, peak \subD{position}\addD{energy} = (1.30\addD{$\,\pm\,$0.03})\,meV and FWHM = (0.4\addD{$\,\pm\,$0.1})\,meV.
			For the higher-energy INS mode, peak \subD{position}\addD{energy} = (1.76\addD{$\,\pm\,$0.03})\,meV and FWHM = (0.4\addD{$\,\pm\,$0.1})\,meV.
			}
	\label{fig:H_Cut}
\end{figure}

The INS data along the H-reciprocal-lattice direction, shown in Fig. \ref{fig:INS_Data}, is a 2D cut where neutron counts have been integrated within the interval [--2.5,2.5] along L, and [--0.2,0.2] along K. \add{The data within the volume along L is relatively non-dispersive, and the intensity is distributed over a broadened energy range}, as shown in Section VI of the supplemental material \cite{supplemental}. The measurements probe the low-energy region of atacamite's spin-wave excitations with an upper-energy threshold of 2.5\,meV.

\begin{figure}[b!]
	\centering
	\includegraphics[width=0.49\textwidth,trim={0cm 0cm 0cm 0cm},clip]{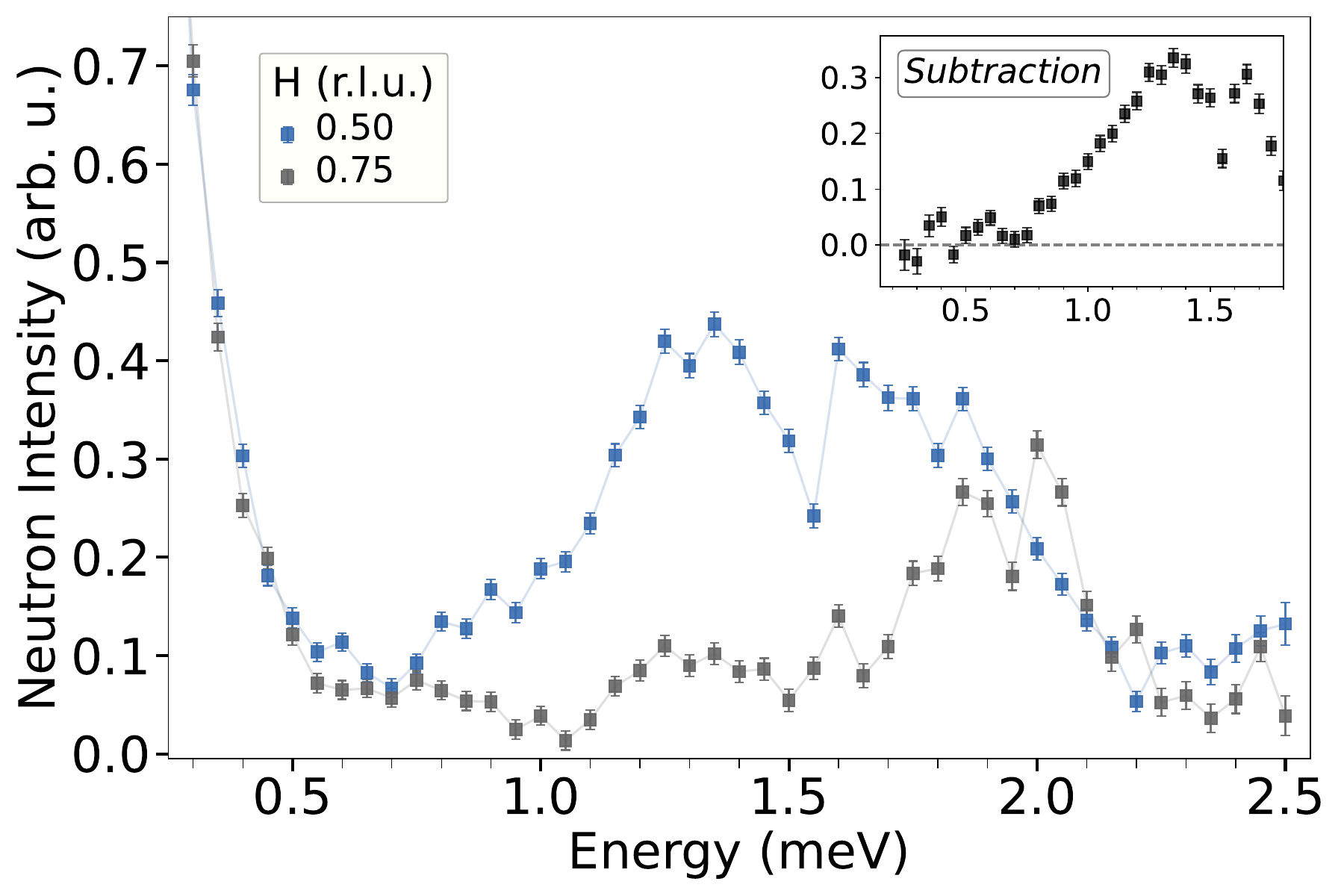}
	\caption{\subD{A 1D cut of the INS data taken at H = 0.50\,r.l.u. overlaid with a 1D cut taken at  H = 0.75\,r.l.u., both with an integration width of Q = 0.04\,r.l.u. along H.}
	\addD{Constant-$Q$ INS spectra at (0.50,0,0) and at (0.75,0,0), both with an integration width of $Q$ = 0.04\,r.l.u..} 
	Neutron counts are within error of each other from below 0.75\,meV, indicating that this is the minimum of the gap. The inset shows a subtraction of the\subD{1D cut of the INS data taken at H = 0.75\,r.l.u.}
	\addD{constant-$Q$ spectrum at (0.75,0,0)} from \subD{the 1D cut taken at  H = 0.50\,r.l.u.}\addD{the spectrum at (0.50,0,0)}. A fiducial line is drawn along zero counts to guide the eye.
			}
	\label{fig:gap_determination}
\end{figure}

Along \subD{the H direction}\addD{(H,0,0)}, two distinct, dispersive, energy-offset spin-wave modes are observed. A linear energy cut at H = 0.5\,r.l.u. is shown in Fig. \ref{fig:H_Cut} and is taken with the same integration widths in L and K as the H-dependent spectra in Fig. \ref{fig:INS_Data}, with an integration width of 0.04\,r.l.u. in H. These data reveal two strong peaks in intensity -- one at 1.7\,meV and one at 1.3\,meV. The lower-energy 1.3\,meV peak is broad, and the neutron count doesn't reach the background level \addB{at the magnetic zone center} until approximately 0.75\,meV, \addB{as shown in Fig. \ref{fig:gap_determination}}
(see Fig. S10 \cite{supplemental} for more details on the gap quantification). Atacamite thus has an energy gap of at least 0.75\,meV along the H\add{00} direction, \add{but, due to instrumental broadening, could be as high as the peak center of $\sim\,1.3$\,meV, which would be compatible with the gap extracted from heat capacity measurements \addD{of approximately 1\,meV} \cite{heinzeAtacamiteCu22025}.}

LSWT calculations performed using the five \subD{largest}\addD{strongest} exchange interactions yielded by \add{previous} DFT results \cite{Heinze_2021}, which includes both the sawtooth and zigzag chains, fail to accurately reproduce \add{all characteristics} of these INS measurements. Specifically, spin-wave excitations appear at higher energies than are observed in the INS measurements, the asymmetry of the mode intensities do not match, and the lower-energy mode is not gapped (for further LSWT details for the full-DFT coupling network, see Figs. S5 and S6 \cite{supplemental}).

\begin{figure}[b!]
	\centering
	\includegraphics[width=0.9\linewidth,trim={0cm 0cm 0cm 0cm},clip]{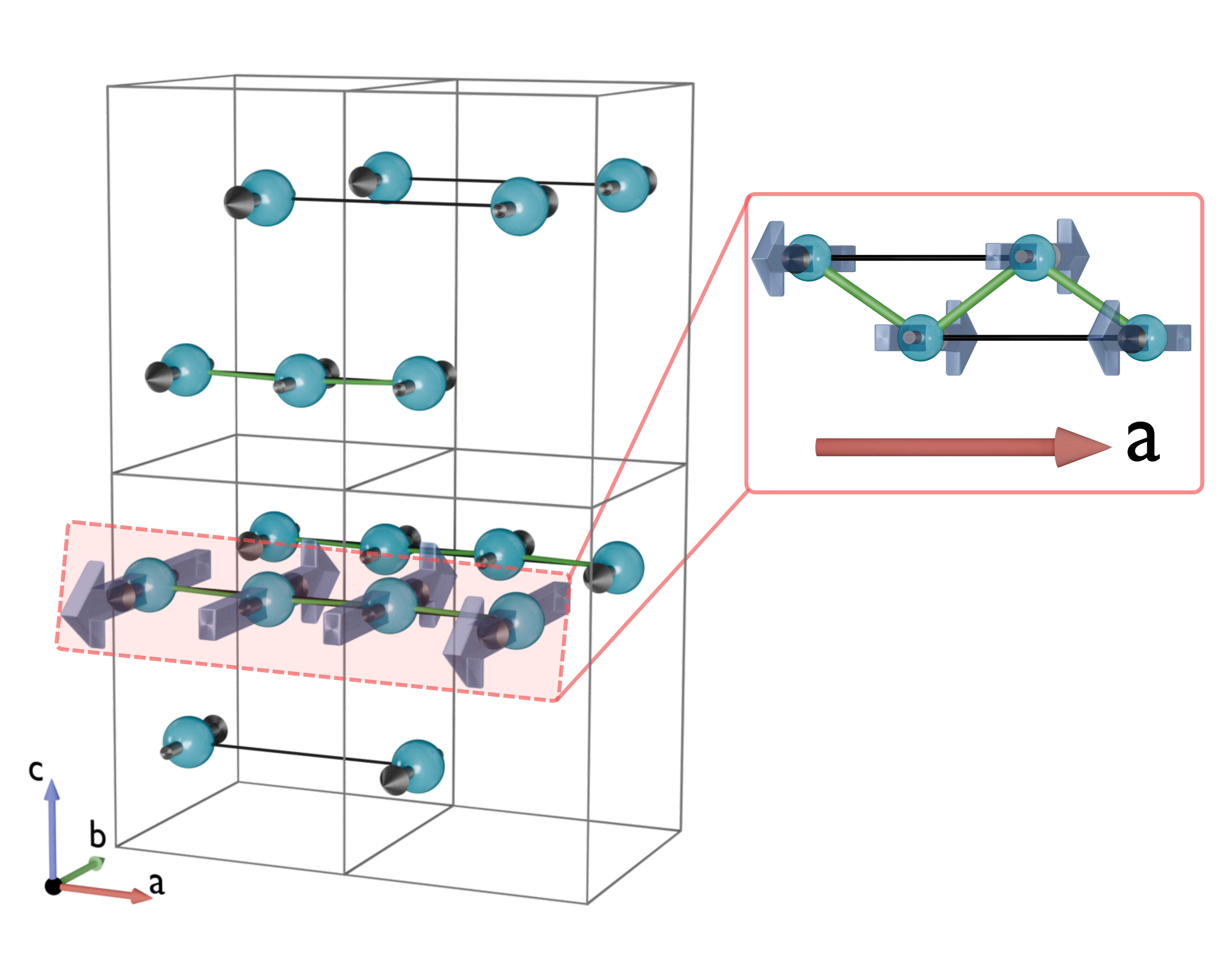}

	\caption{ 
		Atacamite's magnetic structure \cite{Heinze_2021} and isotropic exchange interactions for the zigzag-chain model.
		To the right is the model shown for four Cu(2) atoms with exaggerated angles for clarity.
		Additionally, the staggered mean field components at each spin site \addB{for one spin chain} are represented by transparent blue arrows pointing in the $b$ direction, used to effectively model the residual coupling of the full 3D exchange network.
		}
	\label{fig:Zigzag_Structures}
\end{figure}

These shortcomings demonstrate that the full-DFT model does not adequately capture the magnetism governing atacamite's low-energy excitations within a semiclassical approach.
The discrepancies in the full-DFT model may stem from \subD{a}the simplified treatment of quantum effects in LSWT, and could be alleviated in a fully quantum-mechanical treatment. Unfortunately, frustration and the three-dimensional nature of the full model make such treatment unfeasible. As an alternative, for a given energy scale, we can construct an effective model, which treats a subset of relevant exchanges exactly, while other exchanges enter the model in a form of an effective field \cite{Steven_1996, Aligia_2000, Bursill_1995,Kumar_2015}.
In atacamite, we achieve this by a mean-field treatment of the antiferromagnetic zigzag chain model along the $a$ axis using the Hamiltonian
\begin{equation}
	\mathcal{H}  = \sum_{i=1}^{N}(J_1\mathbf{S}_i\cdot \mathbf{S}_{i+1} + J_{11b}\mathbf{S}_i\cdot \mathbf{S}_{i+2} + \sqrt{2}h\, \textrm{sin}[\frac{\pi}{4}(2i-1)] \mathbf{S}_i^y ).
	\label{eq:atacamite_hamiltonian}
\end{equation}
Here, $J_1$ is the nearest-neighbor interaction \add{along $a$}, $J_{11b}$ is the next-nearest-neighbor interaction for the Cu\add{(2)} spins sites as defined in Fig. \ref{fig:DFT_Structures}, and $h$ is an effective staggered mean-field term which mimics the coupling to Cu(1) spins that are part of the sawtooth chains. The 4-fold periodicity of the staggered field has been constructed from observing the same periodicity in the \add{ordering of the} Cu(2) spin chain itself. This model is represented in Fig. \ref{fig:Zigzag_Structures}.

The parameters of this model have been refined through \subD{least-squares fitting of LSWT calculations to the INS data}\addD{a particle swarm algorithm comparing LSWT calculations to the INS data to find the global minimum}. \add{The} refinement arrives at $J_1$ =  0.90\,meV, $J_{11b}$ = 1.91\,meV, and $h$ = 0.255\,meV, giving a characteristic $J_{11b}$/$J_1$ ratio of 2.12. Classical LSWT calculations of the antiferromagnet\add{ic} zigzag chain with ratios above $\sim$\,1 are qualitatively similar to exact diagonalization calculations (see Fig. S11 in the supplemental information \cite{supplemental}), suggesting quantum effects do not preclude semiclassical calculations from adequately characterizing the excitation spectrum in this system.

The refined LSWT scattering function calculation is shown in Fig. \ref{fig:aniso_zigzag}. The result shows a remarkable qualitative match with experimental data along \subD{the H direction}\addD{(H,0,0)}, namely the spin-wave amplitudes, the relative intensities of the dispersion, asymmetry in the dispersion intensities, the weakening intensity towards the magnetic zone boundary, the spin-wave frequencies, and, crucially, the spin-wave gap. A \subD{1D-cut}\addD{constant-$Q$ cut} of the calculated spin-wave spectra further confirms an excellent match with the experimental peak positions and gap size, as shown in Fig. \ref{fig:H_Cut}.

The zigzag-chain model also provides critical insight into atacamite's gapped ground state. Theoretical considerations of the unmodified zigzag chain \cite{Lavarelo_2014, Steven_1996} suggest that a $J_{11b}$/$J_1$ ratio of $\sim\,$0.5 would be required to produce a gap within the measured range of our INS data. Atacamite's much larger refined ratio of 2.12 would produce a gap in the order of \addC{only} 0.1\,meV \cite{Steven_1996}. This, and our observations that $h$ directly and independently controls the gap size, suggest that the gap is instead primarily a result of interchain interactions \add{modeled} by the mean-field term. Consequently, interchain frustration, particularly via $J_3$ connecting the transverse chains, must be the dominant mechanism driving the observed gap. This is consistent with two observations. One -- that the spin configurations of the Cu(1) and Cu(2) sublattices are \add{near-}perpendicular, a sign that the zigzag chain has adopted a compromised non-colinear configuration with respect to $J_3$ to minimize frustration. And two -- the Cu(2) spins have four-fold commensurate periodicity.

Without frustration, a zigzag chain is expected to adopt an incommensurate \subD{spiral}\addD{helical} order which is nearly\subD{of} four-fold periodic \cite{Steven_1996}. The fact that the ground state is \textit{exactly} four-fold commensurate implies that \add{weak} quantum fluctuations have selected this spin state through order-by-disorder from a frustration-induced classically degenerate set. This breaks symmetry, explaining the large observed gap \cite{henley_ordering_1989}.

While the LSWT-refined parameters provide a compelling and accurate model of the spin-wave excitations, they differ significantly from the initial DFT-calculated exchange values. The nominal DFT values were calculated using the generalized gradient approximation (GGA) method \cite{PBE96} with a Coulomb repulsion $U$-value of 8.24\,\add{eV} \cite{Heinze_2021}. These are $J_1$ =  0.11\,meV and $J_{11b}$ = 1.35\,meV, with a $J_{11b}$/$J_1$ ratio of 12.27, which fail to provide agreement with INS measurements \addC{\cite{supplemental}}.

However, the $U$ value has a drastic influence on the value and even the sign of $J_1$ \cite{Heinze_2021}, and hence also the $J_{11b}$/$J_1$ ratio, which plays the central role for the ground state. To inspect this difference closely, we performed new DFT calculations using the full-potential code FPLO version 22 \cite{FPLO}. We used the GGA supplemented with the Coulomb repulsion $U$ which we varied in the range between 7 and 9\,eV, and the Hund exchange of 1\,eV. Further details of the calculations are provided in Section IV the supplemental information \cite{supplemental}. The GGA+$U$ energies were mapped onto a classical Heisenberg model. In this way, we found that the LSWT-determined ratio $J_{11b}/J_1$ is reproduced by the interpolated values of $J_{11b}$ = 1.40 meV and $J_1$ = 0.66 meV at $U=7.75$ eV. The exchange-coupling values for initial DFT calculations from the literature, our new DFT calculations, and the fitting results for the zigzag chain are shown in Table \ref{tbl:models}.

\begin{table*}
	\caption{Exchange interactions for previous DFT calculations ($U$ = 8.24\,eV) \cite{Heinze_2021}, our updated DFT calculations ($U$ = 7.75\,eV), and the proposed zigzag-chain model whose values come from fitting LSWT calculations to the experimental INS data. Dashes (--) represent non-applicable values.
		A positive $J$-value corresponds to an antiferromagnetic coupling. An effective staggered mean field along the $b$ axis is represented by $h$.
		\addC{Note that the `New DFT' calculations do not include $J_4$ as part of the Hamiltonian (more details are in the supplemental \cite{supplemental}).} 
		}
	\begin{tabular}{crrrr}
		\hline
		Exchange  &\hspace{1.5em} Previous DFT \cite{Heinze_2021}  &\hspace{2em} New DFT   &\hspace{2em} Zigzag-Model LSWT Refinement \\
		 label	 &\hspace{1.5em} 	 		(meV)   &\hspace{2em}		 (meV)	 &\hspace{2em} (meV) \\ 
		\hline
		$J_1$  &	0.11	&	 0.66  			&  0.90 \\
		$J_2$  &	-0.83	&	 -0.83		 	& -- \\
		$J_3$  & 	8.79	&	9.39			& -- \\
		$J_4$  & 	28.97	&	-- 				& -- \\
		$J_{11b}$  & 1.35	&	1.40 			& 1.91 \\
		$h$    & 	--		&	--    			& 0.26 \\
		\hline 
	\end{tabular}
	\label{tbl:models}
\end{table*}

\section{\label{Discussion}Discussion}



\add{Our results have revealed the critical nature of $J_1$ in modeling atacamite, and INS has allowed us to refine its highly $U$-sensitive value.} These results, however, do not \add{directly} cast doubt on the validity of the previous DFT \add{approach}; the previous calculations correspond to the energy scale of the Curie-Weiss temperature \cite{Heinze_2021}, their dominant couplings agree with the Goodenough-Kanamori-Anderson rule \cite{Goodenough_1955,Kanamori_1959,Rocquefelte_2012} by association with atacamite's \subD{largest}\addD{strongest} Cu-O-Cu superexchange pathways, and they contain the very zigzag chain geometry used for our model.

Instead we make three key observations: Firstly, the zigzag chain plays an important role in atacamite's magnetism not previously explored. The necessity of the zigzag chain to reproduce the ordered-phase spin-wave dispersion at 1.5\,K indicates that the zigzag-chain couplings are essential to the magnetic ordering behavior in atacamite. Atacamite's measured ground state behaves as a zigzag chain in a weak staggered field. It is therefore not enough to simply describe atacamite as a one-dimensional sawtooth chain system when discussing its long-range ordered phase. 

Secondly, the energy scales between the zigzag and sawtooth chain are \add{effectively} separated \addD{by one order of magnitude}, reflecting different roles in atacamite's \add{spin} dynamics. Although all \textit{\addC{isotropic}} couplings can be deduced simultaneously through DFT, \add{low-energy} spin-wave calculations performed with the full coupling scheme \add{are unable to fully reproduce} INS observations. Meanwhile, at temperatures \add{far} above T$_\text{N}$ = \subC{8.4\,K}\addC{8.9\,K}, the Curie-Weiss temperature \addD{of $\sim$130\,K} derived from inverse susceptibility \cite{Heinze_2018} aligns with the dominant energy scale of the sawtooth chain \cite{Heinze_2021}. \add{Additionally, magnetic order is suppressed near the energy scale of the zigzag chain \cite{heinzeAtacamiteCu22025}, suggesting it governs atacamite's ordering dynamics.}

This \add{effective} energy scale separation may be explained by frustration effects in atacamite. The refined mean-field term, $h = 0.26$\,meV, is significantly smaller than the dominant interchain coupling predicted by DFT ($\sim$ 9\,meV), indicating that frustration suppresses the effective interchain coupling. Driven by $J_3$, this frustration not only inhibits long-range order by competing within the dominant sawtooth chain but also manifests itself through the order-by-disorder effects discussed earlier. As a result, atacamite's magnetic ground state is more sensitive to fluctuations, which both enhances the importance of subtle Hamiltonian terms, such as anisotropic interactions, and effectively separates the system into two distinct 1D coupling schemes. These same features also make it particularly challenging to construct a complete and precise model based on direct-exchange DFT parameters alone. Small inaccuracies in subtle interactions can become amplified in a fluctuation-influenced system, where semiclassical calculations may respond sensitively to such details.


Lastly, $J_1$ is extremely sensitive to $U$, which means a highly-precise determination of $U$ is crucial for a model of atacamite. Through LSWT refinement against INS measurements, we have provided a significant and thorough refinement of the $U$ value required to accurately describe atacamite in the low-temperature regime associated with long-range order.

\section{\label{Conclusion}Conclusions}

LSWT calculations and INS measurements have revealed key insights into atacamite's magnetism in its long-range ordered phase. The mean-field-modified zigzag-chain model accurately reproduces the periodicity, intensity asymmetry, dispersion, and spin-wave gap observed in the INS data. The effective mean-field term $h$ estimates the frustrated interchain interaction strength and provides insight into the spin-wave gap through an order-by-disorder mechanism. 


In contrast, while prior DFT calculations for the dominant direct exchange couplings are consistent with high-temperature experimental results, they fail to account for the low-energy spin dynamics below the ordering temperature in a semiclassical LSWT treatment. This suggests that \subD{the full isotropic-exchange DFT model may be missing the subtleties required to capture the low-energy magnetic dynamics in a frustrated magnet.}\addD{the DFT+$U$ approach is unable to describe the magnetism across different energy scales with a single $U$ value in this complex and frustrated magnetic system.} Frustration also enhances quantum effects, suppressing weaker interactions such as interchain couplings. This has led to an effective energy scale separation between two distinct 1D-chain models. Nevertheless, by employing an effective model of a field-moderated zigzag chain, we successfully capture the essential features of the experimental dispersion relations and provide a reliable description of atacamite's low-energy magnetism.
\addC{This way, we have uncovered the frustrated interchain exchange in atacamite leading to long-range magnetic order below its Néel temperature, and where the magnetic state is selected through order-by-disorder. In the light of the recent experimental results in high magnetic fields \cite{heinzeAtacamiteCu22025}, it seems to be this frustrated interchain coupling network which is effectively broken up when the Cu(2) spins are fully field polarized in magnetic fields beyond the quantum critical point. Numerical results here supported the scenario of an (effective) dimensional reduction of the spin system upon which long-range magnetic order is suppressed.}

\addB{Through establishing the exchange interactions outside the sawtooth chain in more detail, an avenue is now opened for building a clearer picture of the field-induced transitions which depend on the spin dynamics connected to these subtler interactions.}

\section*{\label{Acknowledgments}Acknowledgments}
This research used the Pelican time-of-flight spectrometer for proposal number P8226 at ACNS, a part of ANSTO, the Australian Nuclear Science and Technology Organisation. J. L. Allen would like to thank AINSE Limited for providing financial assistance (Award - PGRA) to enable this research. O.J. and S.N. were supported by the German Forschungsgemeinschaft (DFG, German Research Foundation) through SFB 1143 (Project ID 247310070). The authors acknowledge and thank A. U. B. Wolter from The Leibniz Institute for Solid State and Materials Research for her extensive contributions to atacamite research, and for constructive discussions while preparing the manuscript. Finally, we thank Ulrike Nitzsche for technical assistance.

\bibliography{ata_2022}

\end{document}



\title{The Magnetic Ground State of Atacamite Cu$_2$Cl(OH)$_3$: The Crucial Role of Frustrated Zigzag Chains Revealed by Inelastic Neutron Scattering -- Supplemental Information}



\author{J. L. Allen}
\email[]{jlallen.correspondence@outlook.com}
\affiliation{Institute for Superconducting and Electronic Materials and School of Physics, University of Wollongong, NSW 2522, Australia}

\author{L. St\"odter}%
\affiliation{Institut für Physik der Kondensierten Materie, Technische Universität Braunschweig, D-38106 Braunschweig, Germany}
\affiliation{Jülich Centre for Neutron Science (JCNS) at Heinz Maier-Leibnitz Zentrum (MLZ), Forschungszentrum Jülich GmbH, 85748 Garching, Germany}

\author{R. A. Mole}
\affiliation{Australian Centre for Neutron Scattering, Australian Nuclear Science and Technology Organisation, \\Lucas Heights, NSW 2234, Australia}

\author{S. S\"ullow}
\affiliation{Institut für Physik der Kondensierten Materie, Technische Universität Braunschweig, D-38106 Braunschweig, Germany}

\author{O. Janson}
\affiliation{Institute for Theoretical Solid State Physics, Leibniz Institute for Solid State and Materials Research Dresden, 01069 Dresden, Germany}

\author{S. Nishimoto}
\affiliation{Institute for Theoretical Solid State Physics, Leibniz Institute for Solid State and Materials Research Dresden, 01069 Dresden, Germany}
\affiliation{Department of Physics, Technical University Dresden, 01069 Dresden, Germany}

\author{R. A. Lewis}
\affiliation{Institute for Superconducting and Electronic Materials and School of Physics, University of Wollongong, NSW 2522, Australia}

\author{K. C. Rule}
\email[]{kirrilyr@gmail.com}
\affiliation{Australian Centre for Neutron Scattering, Australian Nuclear Science and Technology Organisation, \\Lucas Heights, NSW 2234, Australia}
\affiliation{Institute for Superconducting and Electronic Materials and School of Physics, University of Wollongong, NSW 2522, Australia}


\date{\today}

\maketitle


\section{\label{Chemical}Chemical Structure of Atacamite}

\begin{figure}[h!]
	\centering
	\includegraphics[width=0.75\textwidth,trim={0cm 0cm 0cm 0cm},clip]{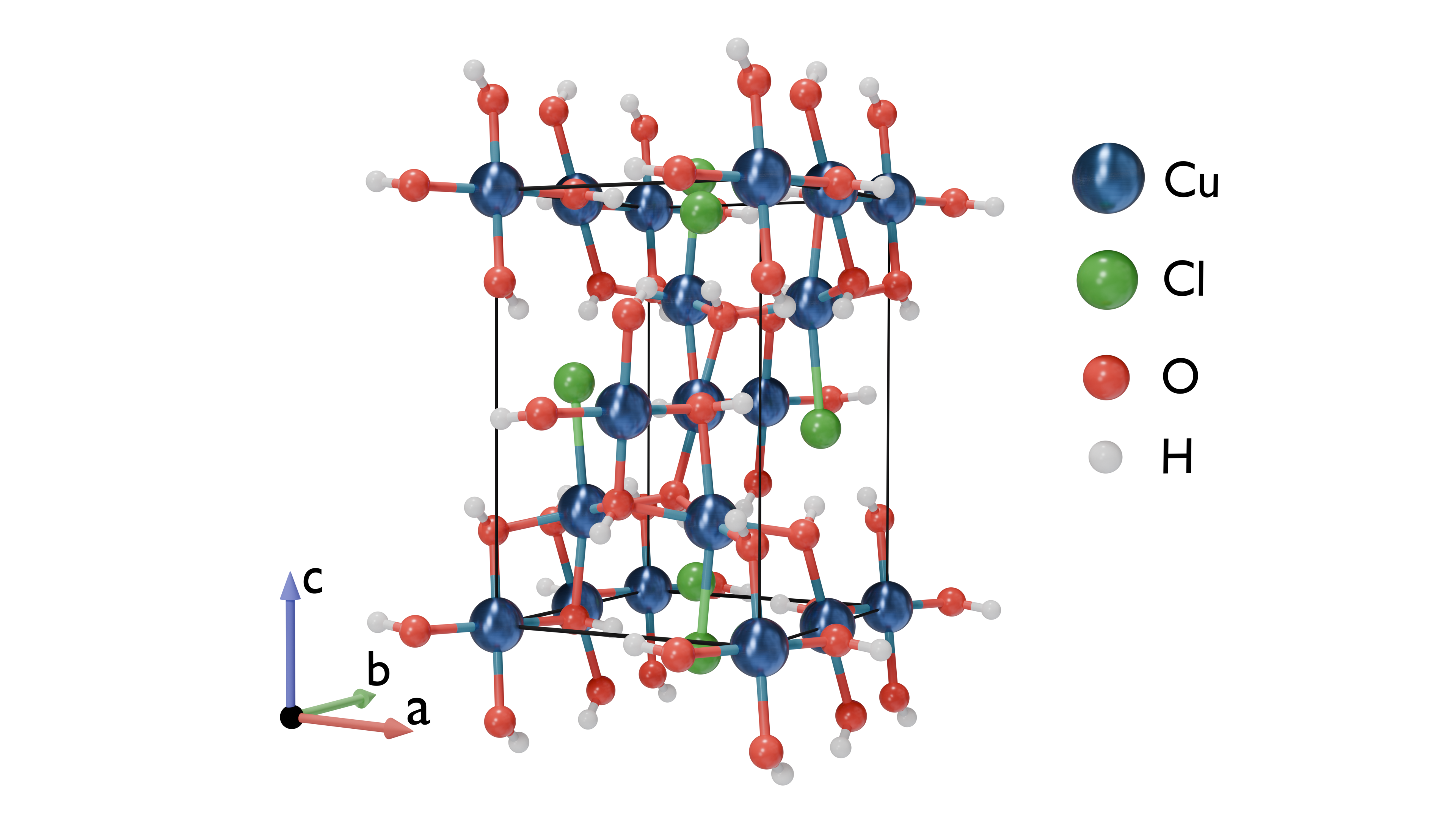}
	\caption{The crystal structure of atacamite determined by X-ray diffraction data \cite{Parise_1986}.}
	\label{fig:Chem_structure}
\end{figure}

\begin{table}[h!]
	\caption{Bond length and angles for Cu-O-Cu bonds in atacamite. The bond in row 3 is has an irregular rectangular shape whose lack of symmetry gives Cu-O-Cu branches of uneven length and angle. Thus two entries for Cu-O bond length and angle are given for each branch. Data is extracted from Ref. \cite{Parise_1986}.}
	\begin{tabular}{clll}
		\hline
		Bond  &\hspace{1em} Axis &\hspace{1em} Cu-O bond &\hspace{1em} Cu-O bond \\
		Type &\hspace{1em} Direction &\hspace{1em} Length (\AA) &\hspace{1em} Angle ($\degree$) \\ 
		\hline
		\vspace{-0.75em}&&&\\
		\chemfig{Cu_1-O-Cu_1}  &\hspace{1em} $b$ &\hspace{1em}  1.940(3) &\hspace{1em} 124.5(3) \\
		&&&\\
		\chemfig{Cu_1-O-Cu_2}  &\hspace{1em}  $a,b,c$  &\hspace{1em} 1.993(3), 2.017(2) &\hspace{1em} 114.74(11) \\
		&&&\\
		\chemfig{Cu_1(-[:20]O-[:-20]{\color{white}Cu_2})(-[:-20]O-[:20]Cu_2)}  &\hspace{1em}  $a,b,c$ &\hspace{1em} 1.940(3), 2.010(3),   &\hspace{1em} 92.30(10), \\
		&&\hspace{1em} 2.017(2), 2.358(3)&\hspace{1em} 101.23(11)\\
		&&&\\
		\chemfig{Cu_2(-[:20]O-[:-20]{\color{white}Cu_2})(-[:-20]O-[:20]Cu_2)}  &\hspace{1em}  $a$ &\hspace{1em} 2.010(3), 1.993(3)  &\hspace{1em} 97.81(12) \\
		\hline 
	\end{tabular}
	\label{tbl:bonds}
\end{table}

Atacamite (Cu$_2$Cl(OH)$_3$) has an orthorhombic $Pnma$ crystal structure with $a=6.030(2)$ \AA, $b=6.865(2)$ \AA, and $c=9.120(2)$ \AA \cite{Parise_1986}. The structure consists of two inequivalent Cu sites which make 3 unique Cu-O-Cu bonds whose bond lengths and bond angles are shown in Table \ref{tbl:bonds}. The exchange interactions associated with each bond, in descending order, are $J_4$, $J_3$, $J_2$, and $J_1$.

Generally speaking, the Goodenough-Kanamori-Anderson rule\addC{s} \cite{Goodenough_1955,Kanamori_1959} states that for cation-anion-cation bonds, $90\degree$ bonds correspond to ferromagnetic superexchange interactions while $180\degree$ bonds correspond to antiferromagnetic superexchange interactions. For Cu-O-Cu bonds in antiferromagnetic materials specifically, antiferromagnetic superexchange interactions are weak in magnitude at $90\degree$ and become increasingly stronger towards $180\degree$\cite{Rocquefelte_2012}. The significantly larger bond angle of $124.4\degree$ for the bond associated with \addC{$J_4$} along the $b$ axis is therefore consistent with this model assigning that bond the largest antiferromagnetic exchange interaction.
\newpage
\clearpage

\section{\label{Full Structural Representation}Full Representation of Magnetic Structures}

\begin{figure}[h]
	\centering
	\includegraphics[width=0.9\textwidth,trim={0cm 0cm 0cm 0cm},clip]{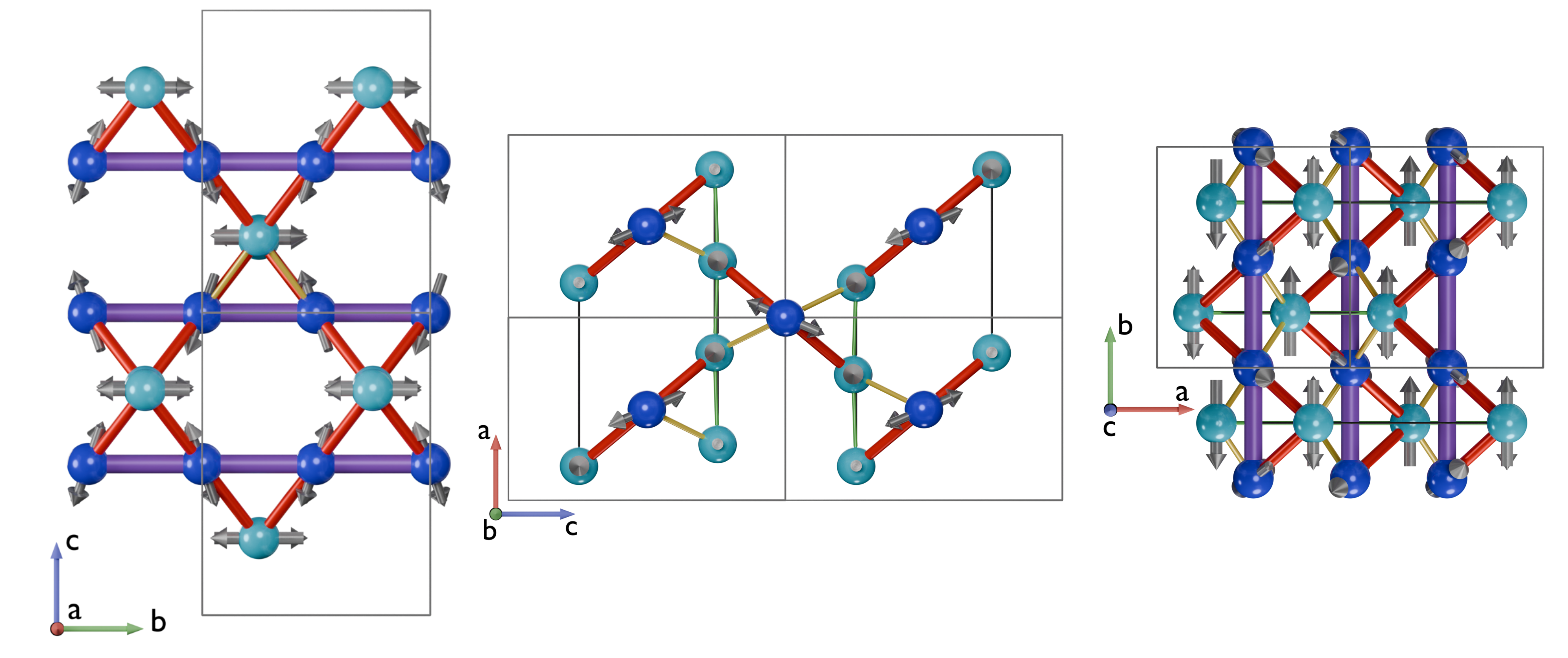}
	\caption{From left to right, the full-DFT model viewed along the \textit{a} axis, the
		\textit{b} axis, and the \textit{c} axis.}
	\label{fig:dft_full}
\end{figure}
\begin{figure}[h]
	\centering
	\includegraphics[width=0.9\textwidth,trim={0cm 0cm 0cm 0cm},clip]{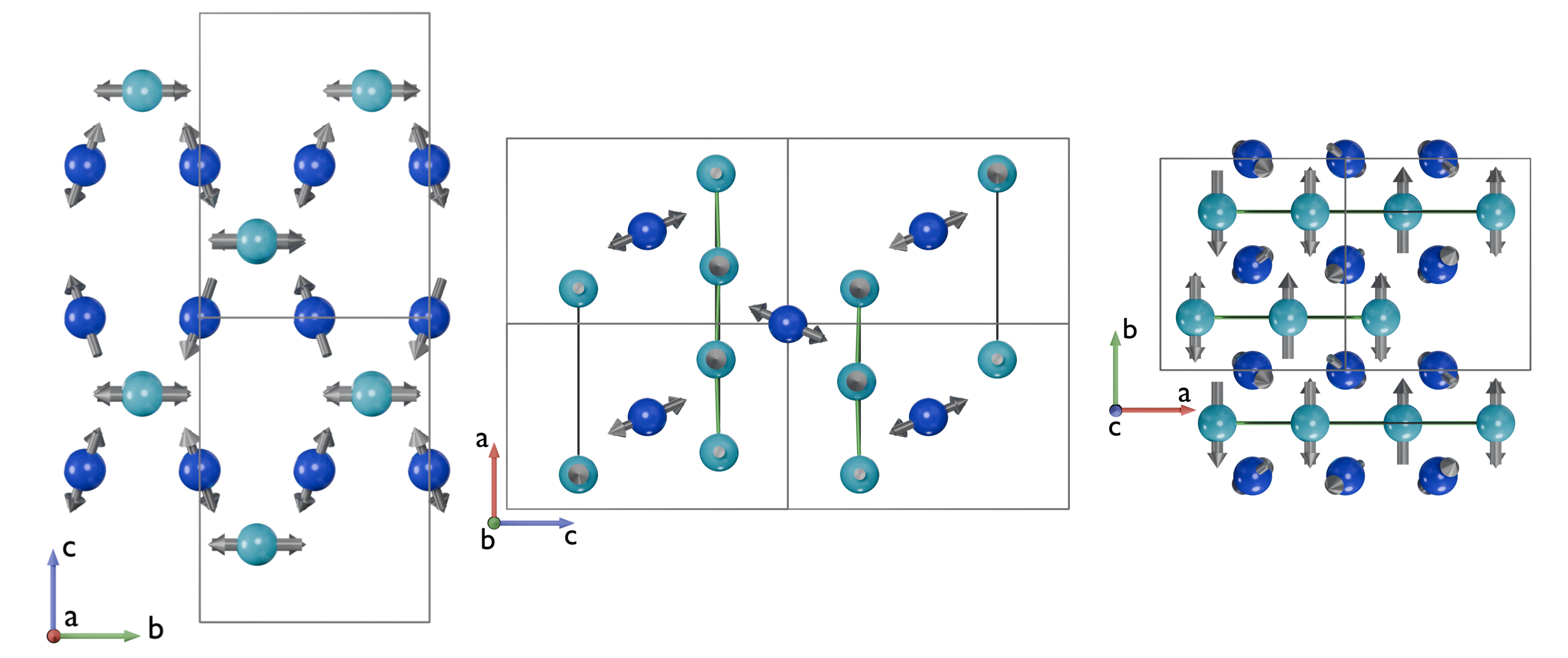}
	\caption{From left to right, the zigzag-chain model viewed along the \textit{a} axis, the
		\textit{b} axis, and the \textit{c} axis.}
	\label{fig:zigzag_full}
\end{figure}

\addC{Views along each principle axis in the $2\times1\times2$ magnetic supercell packing diagram are provided here to help clarify the structural geometry of the magnetic exchange networks in the full-DFT model and the zigzag-chain model. 
Fig. \ref{fig:dft_full} presents the three principle-axis views full-DFT magnetic model, and Fig. \ref{fig:zigzag_full} shows a similar arrangement for the zigzag-chain magnetic model.}

\newpage
\clearpage

\section{\label{INS}Full Set of Inelastic Neutron Scattering Data}

\begin{figure}[h]
	\centering
	\begin{subfigure}[b]{0.32\textwidth}
		\centering
		\includegraphics[width=1\textwidth,trim={0cm 0cm 0.3cm 0cm},clip]{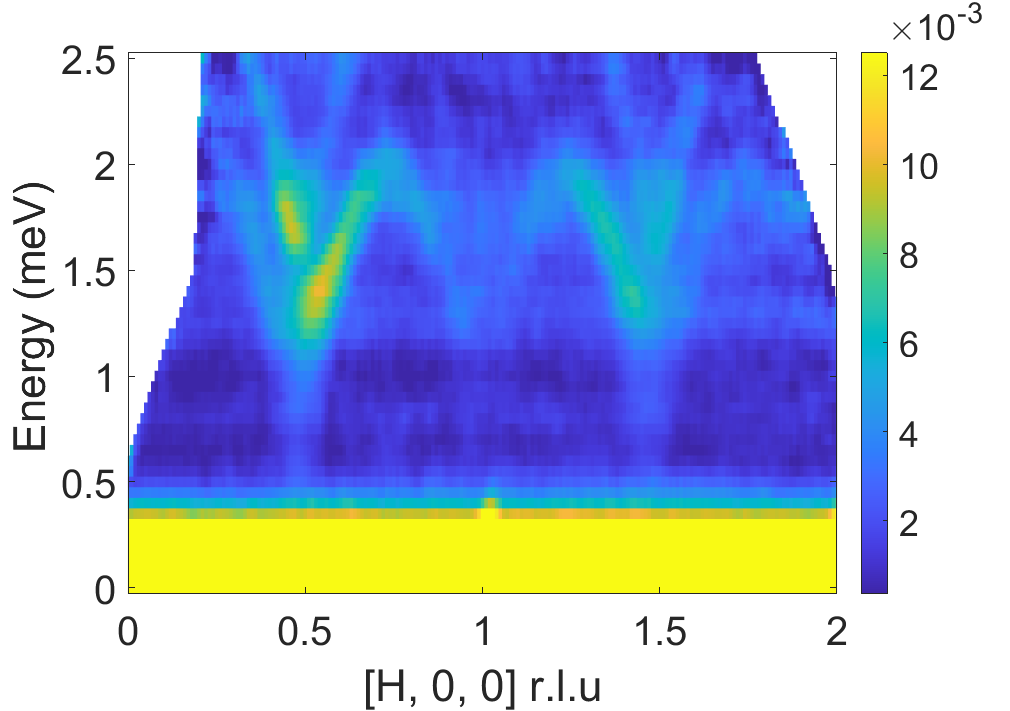}
		\caption{}
		\label{fig:H_data}
	\end{subfigure}
	\begin{subfigure}[b]{0.32\textwidth}
		~
		\centering
		\includegraphics[width=1\textwidth,trim={0cm 0cm 0.3cm 0cm},clip]{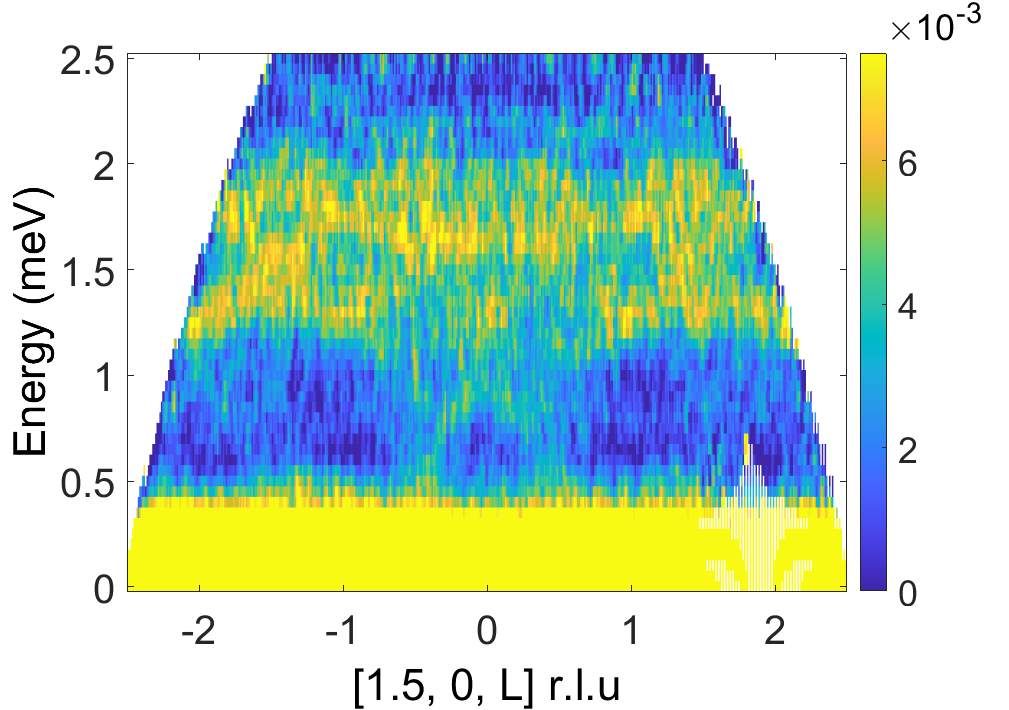}
		\caption{}
		\label{fig:L_data}
	\end{subfigure}
	\begin{subfigure}[b]{0.32\textwidth}
		~
		\centering
		\includegraphics[width=1\textwidth,trim={0cm 0cm 0.3cm 0cm},clip]{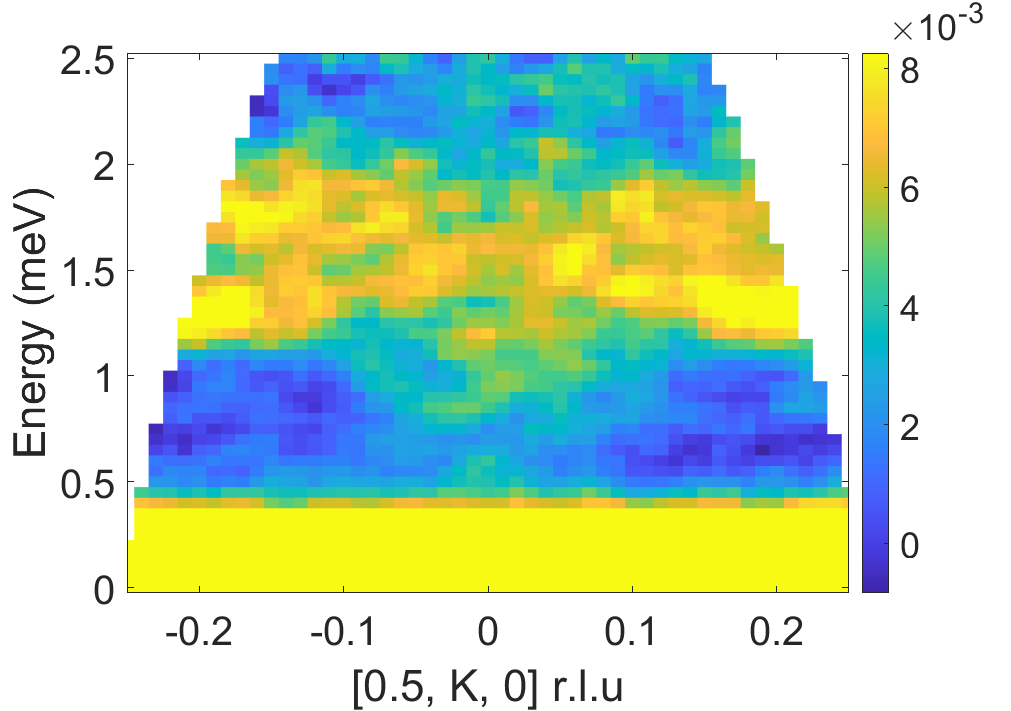}
		\caption{}
		\label{fig:K_data}
\end{subfigure}
	\caption{Experimental INS data of (co-aligned) single-crystalline atacamite measured on Pelican at ANSTO. (a) Strong dispersion can be observed in the H direction with two separate nested modes. (b) INS spectra measured along the L direction. (c) INS spectra measured along a narrow out-of-plane range in the K direction. The colorbars represent neutron count intensity.}
	\label{fig:INS_Data}
\end{figure}

The dynamics of the spin-wave modes along \subD{the L direction}\addD{(1.5,0,L)}, shown in Fig. \ref{fig:L_data} are harder to distinguish due to the poorer resolution (which is itself due to a smaller area of reciprocal space coverage compared to the data in the \subD{H}\addD{(H,0,0)} direction). The observed pattern is consistent with multiple sinusoidal spin-wave modes between 1 and 2\,meV, with additional modes emerging from the [$\frac{3}{2}$ 0 $\frac{1}{2}$] magnetic zone center.
\subD{In combination with DFT results suggesting no direct exchange interactions along the $c$ axis, this makes relating the spin-wave dynamics to the magnetic structure along this direction difficult to analyze.}

\addD{In terms of the DFT-calculated interactions, the exchange couplings $J_2$ and $J_3$ are interchain couplings along the $c$ axis. However, they are not entirely along $c$; these are off-axis couplings, containing components along the $a$ and $b$ axes. Thus, modeling the spin-waves along $c$ would increase the degrees of freedom of the zigzag-chain model. The complexity of the $c$-axis couplings, and the relatively low-quality of the data along (1.5,0,L), makes relating the spin-wave dynamics to the magnetic structure along this direction difficult to analyze.}

The out-of-plane magnon scattering seen along \subD{the K direction}\addD{(0.5,K,0)}, shown in Fig. \ref{fig:K_data} is over a limited range of --0.2 to 0.2\,r.l.u. Despite this, dispersive modes are seen, with likely two branches crossing at 0\,r.l.u. A wider range over a full magnetic zone range is required for meaningful comparison to modeling.
\newpage
\clearpage

\section{Details of DFT calculations}

For consistency with the Ref. \cite{Heinze_2021}, we used the same structural input as in that publication: unit cell parameters and atomic coordinates of Cu, Cl, and O atoms from Ref. \cite{Zheng_2005_B}; for the H atoms, we used the optimized coordinates from Ref. \cite{Heinze_2021}. The resulting cell was doubled along the $a$ axis and recast in the space group $Pmn2_1$ (31), which features six Cu Wyckoff positions and in which the $a$ and $b$ axes are interchanged. Atomic coordinates are given in Table~\ref{tab:coords}.

\begin{table}[h]
\caption{\label{tab:coords}Atomic coordinates within the $Pmn2_1$ (31) supercell ($a$ = 6.86383\,\r{A}, $b$ = 12.05594,\r{A}, $c$ = 9.11562,\r{A}) used for total energy GGA+$U$ calculations.}
\begin{ruledtabular}
\begin{tabular}{rrddd}
atom & Wyckoff position & x/a & y/b & z/c \\ \hline
Cu1 & $4b$ & 0.75    & 0.125   & 0       \\
Cu2 & $4b$ & 0.75    & 0.625   & 0       \\
Cu3 & $2a$ & 0       & 0.028715 & 0.25603 \\
Cu4 & $2a$ & 0       & 0.778715 & 0.24397 \\
Cu5 & $2a$ & 0       & 0.528715 & 0.25603 \\
Cu6 & $2a$ & 0       & 0.278715 & 0.24397 \\
Cl1 & $2a$ & 0       & 0.04570 & 0.55650 \\
Cl2 & $2a$ & 0       & 0.54570 & 0.55650 \\
Cl3 & $2a$ & 0       & 0.79570 & -0.05650 \\
Cl4 & $2a$ & 0       & 0.29570 & -0.05650 \\
O1 & $2a$ & 0       & 0.05010 & -0.00686 \\
O2 & $2a$ & 0       & 0.55010 & -0.00686 \\
O3 & $2a$ & 0       & 0.30010 & 0.50686 \\
O4 & $2a$ & 0       & 0.80010 & 0.50686 \\
O5 & $4b$ & 0.81360 & -0.09470 & 0.28750 \\
O6 & $4b$ & 0.81360 & 0.40530 & 0.28750 \\
O7 & $4b$ & 0.68640 & 0.84470 & 0.71250 \\
O8 & $4b$ & 0.68640 & 0.34470 & 0.71250 \\
H1 & $2a$ & 0       & 0.22120 & 0.53880 \\
H2 & $2a$ & 0       & 0.72120 & 0.53880 \\
H3 & $2a$ & 0       & -0.02880 & -0.03880 \\
H4 & $2a$ & 0       & 0.47120 & -0.03880 \\
H5 & $4b$ & 0.30230 & -0.08685 & 0.22140 \\
H6 & $4b$ & 0.30230 & 0.41315 & 0.22140 \\
H7 & $4b$ & 0.19770 & 0.83685 & 0.77860 \\
H8 & $4b$ & 0.19770 & 0.33685 & 0.77860 \\ 
\end{tabular}
\end{ruledtabular}
\end{table}

Total energy GGA+$U$ calculations were done on a mesh of $4\times{}2\times3$ $k$-points. We used the fully-localized limit for the double-counting correction and varied the Coulomb repulsion $U$ in the range between 7 and 9\,eV by keeping the Hund exchange fixed to 1\,eV. The resulting total energies were mapped onto a classical Heisenberg model with six exchanges: $J_1$, $J_2$, $J_3$, $J_{11b}$ $J_{13}$, and $J_{14}$. Note that the leading magnetic exchange $J_4$ is not part of this Hamiltonian: the respective spins in the supercell are equivalent by symmetry, and thus yield a constant energy contribution in each configuration. For each magnetic configuration $i$, the total energy $E_i$ can be expressed as
\begin{equation}
  \label{eq:heisenberg}
  E_i = E_0 + \sum_{k}{c_{ik} J_k},
\end{equation}
where $E_0$ is a constant and $k$ runs over the subset of relevant exchanges (1, 2, 3, $11b$, 13, 14). The coefficients $c_{ik}$ are listed in Table~\ref{tab:coeffs}.

\begin{table}[h]
\caption{\label{tab:coeffs}Coefficients $c_{ik}$ in the classical Heisenberg model Eq.~(\ref{eq:heisenberg}) describing different magnetic configurations $i$. For coordinates of Cu1--Cu6 atoms, see Table~\ref{tab:coords}.}
\begin{ruledtabular}
\begin{tabular}{ddddddrrrrrr}
\multicolumn{6}{c}{$i$ (magnetic configuration)} & \multicolumn{6}{c}{$c_{ik}$ coefficients in the Heisenberg Hamiltonian} \\
\text{Cu1} & \text{Cu2} & \text{Cu3} & \text{Cu4} & \text{Cu5} & \text{Cu6} & $k\!=\!1$ \hspace{3em}& $k\!=\!2$ \hspace{3em}& $k\!=\!3$ \hspace{3em}& $k\!=\!{11b}$ \hspace{3em}& $k\!=\!{13}$ \hspace{0em}& $k\!=\!{14}$ \\ \hline
\downarrow &\downarrow &\downarrow &\downarrow &\downarrow & \downarrow & 2 &  4 &  4 &  2 &  2 &  4 \\
\downarrow &\downarrow &\downarrow &\downarrow &\downarrow &   \uparrow & 0 &  2 &  2 & 0 & 0 &  4 \\
\downarrow &\downarrow &\downarrow &\downarrow &  \uparrow &   \uparrow & 0 & 0 & 0 & -2 &  2 &  4 \\
\downarrow &\downarrow &\downarrow &  \uparrow &\downarrow &   \uparrow & -2 & 0 & 0 &  2 & -2 &  4 \\
\downarrow &\downarrow &\downarrow &  \uparrow &  \uparrow & \downarrow &  0 & 0 & 0 & -2 & -2 &  4 \\
\downarrow &\downarrow &\downarrow &  \uparrow &  \uparrow &   \uparrow & 0 & -2 & -2 & 0 & 0 &  4 \\
\downarrow &\downarrow &  \uparrow &  \uparrow &  \uparrow &   \uparrow & 2 & -4 & -4 &  2 &  2 &  4 \\
\downarrow &  \uparrow &\downarrow &\downarrow &\downarrow & \downarrow &  2 & 0 & 0 &  2 &  2 & 0 \\
\downarrow &  \uparrow &\downarrow &\downarrow &\downarrow &   \uparrow & 0 &  2 & -2 & 0 & 0 & 0 \\
\downarrow &  \uparrow &\downarrow &\downarrow &  \uparrow & \downarrow &  0 &  2 &  2 & 0 & 0 & 0 \\
\downarrow &  \uparrow &\downarrow &\downarrow &  \uparrow &   \uparrow & 0 &  4 & 0 & -2 &  2 & 0 \\
\downarrow &  \uparrow &\downarrow &  \uparrow &\downarrow & \downarrow &  0 & -2 &  2 & 0 & 0 & 0 \\
\downarrow &  \uparrow &\downarrow &  \uparrow &\downarrow &   \uparrow & -2 & 0 & 0 &  2 & -2 & 0 \\
\downarrow &  \uparrow &\downarrow &  \uparrow &  \uparrow & \downarrow &  0 & 0 &  4 & -2 & -2 & 0 \\
\downarrow &  \uparrow &  \uparrow &\downarrow &\downarrow & \downarrow &  0 & -2 & -2 & 0 & 0 & 0 \\
\downarrow &  \uparrow &  \uparrow &\downarrow &\downarrow &   \uparrow & 0 & 0 & -4 & -2 & -2 & 0 \\
\downarrow &  \uparrow &  \uparrow &  \uparrow &\downarrow & \downarrow &  0 & -4 & 0 & -2 &  2 & 0 \\
\end{tabular}
\end{ruledtabular}
\end{table}

\begin{figure}[h]
    \includegraphics[width=\textwidth]{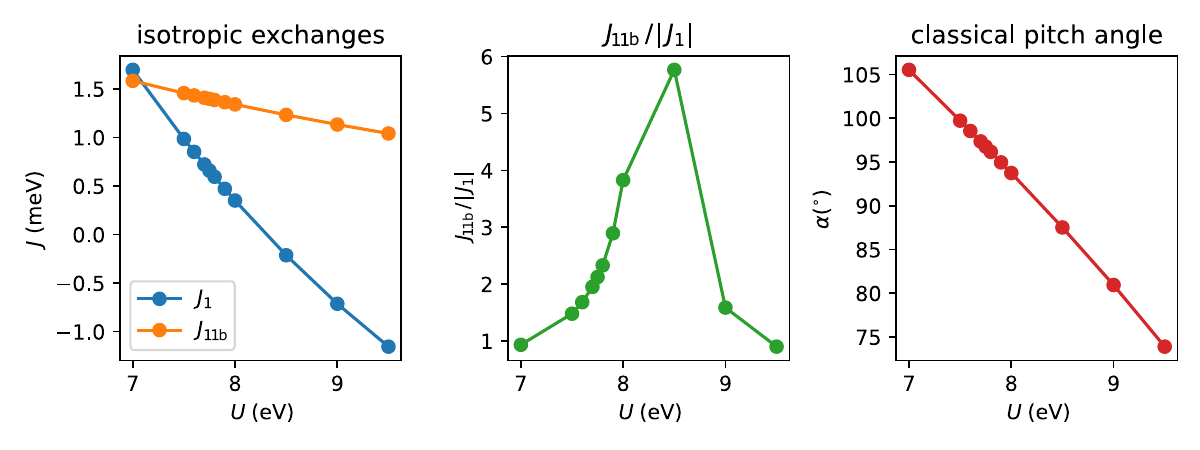}
    \caption{Exchanges $J_1$ and $J_{11b}$ (left), $J_{11b}/|J_1|$ (middle), and the classical pitch angle as a function of $U$ (right).}
	\label{fig:dft}
\end{figure}

Using Eq.~(\ref{eq:heisenberg}) with coefficients from Table~\ref{tab:coeffs} and GGA+$U$ total energies for different values of $U$, we can compute all six exchange integrals with a least-square fitting. These results are provided in Table~\ref{tab:js}. Since our main interests are the $J_1$ and $J_{11b}$ exchanges, we plot their values, their ratio, and the classical pitch angle $\phi=\arccos\left(-\frac{J_1}{4J_{11b}}\right)$ as a function of $U$ in Fig.~\ref{fig:dft}.

\begin{table}[h]
	\caption{Magnetic exchange integrals (meV) obtained from a least-square fit of Eq.~(\ref{eq:heisenberg}) parameterized with GGA+$U$ total energies.}
	\label{tab:js}
	\begin{ruledtabular}
	\begin{tabular}{ddddddd}
	\multicolumn{1}{c}{U \text{(eV)}} & 
	\multicolumn{1}{c}{$J_1$} & 
	\multicolumn{1}{c}{$J_2$} & 
	\multicolumn{1}{c}{$J_3$} & 
	\multicolumn{1}{c}{$J_{11b}$} & 
	\multicolumn{1}{c}{$J_{13}$} & 
	\multicolumn{1}{c}{$J_{14}$} \\ \hline
	7.00 & 1.70 & -0.70 & 10.78 & 1.58 & -0.09 & -0.06 \\ 
	7.50 & 0.98 & -0.79 & 9.83 & 1.46 & -0.06 & -0.05 \\ 
	7.60 & 0.85 & -0.81 & 9.65 & 1.43 & -0.06 & -0.05 \\ 
	7.70 & 0.72 & -0.82 & 9.47 & 1.41 & -0.06 & -0.05 \\ 
	7.75 & 0.66 & -0.83 & 9.39 & 1.40 & -0.06 & -0.05 \\ 
	7.80 & 0.59 & -0.84 & 9.30 & 1.38 & -0.05 & -0.05 \\ 
	7.90 & 0.47 & -0.85 & 9.13 & 1.36 & -0.05 & -0.05 \\ 
	8.00 & 0.35 & -0.86 & 8.96 & 1.34 & -0.05 & -0.04 \\ 
	8.50 & -0.21 & -0.91 & 8.16 & 1.23 & -0.04 & -0.04 \\ 
	9.00 & -0.71 & -0.95 & 7.42 & 1.13 & -0.03 & -0.03 \\ 
	9.50 & -1.15 & -0.97 & 6.75 & 1.04 & -0.02 & -0.02 \\ 
	\end{tabular}
	\end{ruledtabular}
\end{table}

\newpage
\clearpage

\section{LSWT Calculations with Additional Exchange Interactions}

\subsection{Original DFT-Calculated Interactions}

\begin{figure}[h!]
	\centering
	\includegraphics[width=0.5\textwidth,trim={0cm 0cm 0cm 0cm},clip]{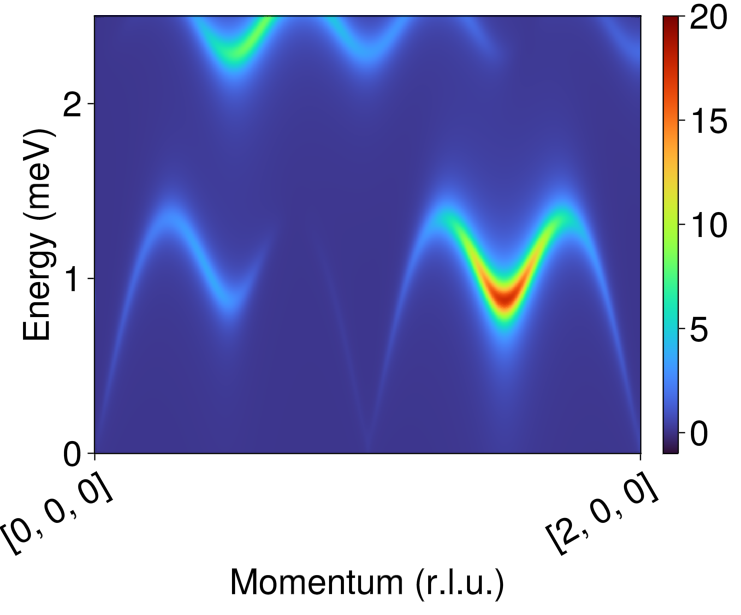}
	\caption{LSWT calculations of the \addD{low-energy} spin-wave modes along \subD{the H direction}\addD{(H,0,0)} in atacamite using the exchange network as given by previous DFT calculations \cite{Heinze_2021}.}
	\label{fig:original_dft_spinw}
\end{figure}

Linear spin-wave theory (LSWT) modeling with the DFT-calculated exchange network for the five largest calculated exchange couplings as shown in Ref. \cite{Heinze_2021} is shown in Fig. \ref{fig:original_dft_spinw}.
The spin configuration was fixed to the experimental results presented in Ref. \cite{Heinze_2021}.
The relevant exchange interactions are as presented in Table \ref{tbl:original_dft_values}.

\begin{table*}[h!]
	\caption{Exchange interaction values for the original `full-DFT' calculations from Ref. \cite{Heinze_2021}. These magnetic exchanges have been used for the Hamiltonian in the calculation shown in Fig. \ref{fig:original_dft_spinw}.}
	\begin{tabular}{cr}
		\hline
		Exchange  &\hspace{1em} Previous DFT \cite{Heinze_2021}   \\
		 label	 &\hspace{1em} 	 		(meV)  \\ 
		\hline
		$J_1$  &	0.11	\\
		$J_2$  & 	-0.83	 \\
		$J_3$  & 	8.79	\\
		$J_4$  & 	28.97	 \\
		$J_{11b}$  &	1.35	 \\
		\hline 
	\end{tabular}
	\label{tbl:original_dft_values}
\end{table*}





\subsection{Conclusions from Additional Exchange Interaction Models}

\subD{Including a broader range of exchange interactions produces LSWT calculations lacking}\addD{The result of calculating the spin-wave spectra of atacamite using the five leading exchange interactions from the original DFT approach \cite{Heinze_2021} lacks} many of the features which represent a match to the INS measurements when compared to the simpler zigzag-chain model. The existence of two dispersive modes and matching periodicity do agree with the INS data in the original DFT-calculated values. However, there is no energy gap in this model. The amplitude of the higher-energy mode in this calculation is also not similar to the real amplitude of this mode from the INS measurements. Additionally, the relative mode intensity does not match the expected asymmetry.

The departure from many of the spin-wave features which match the INS data when applying the full DFT models suggests that the additional DFT-calculated exchange interactions beyond those associated with the zigzag-chain model geometry interestingly are not enough to describe the atacamite system at low temperatures and in the low-energy regime as a complete and explicit model using LSWT.

\addD{Furthermore, spin-wave spectra calculated with LSWT using the `new-DFT' exchange interactions (see Table I in the main manuscript) result in spin-wave modes which share identical characteristics to the `previous-DFT' calculation in Fig. \ref{fig:original_dft_spinw} (bands shift in energy by $\sim$0.1\,meV only). This result indicates that the issues in modeling the low-energy spin-waves modes of atacamite measured by INS are not simply a matter of applying an incorrect $U$ value, and instead, the effective zigzag-chain model is required to capture the main features of atacamite's INS-measured spin waves.}

\newpage
\clearpage

\section{\label{1D INS}L Dependence of Inelastic Neutron Scattering Data}

\begin{figure}[h]
	\centering
	\begin{subfigure}[b]{0.32\textwidth}
		\centering
		\includegraphics[width=1\textwidth,trim={0cm 0cm 0cm 0cm},clip]{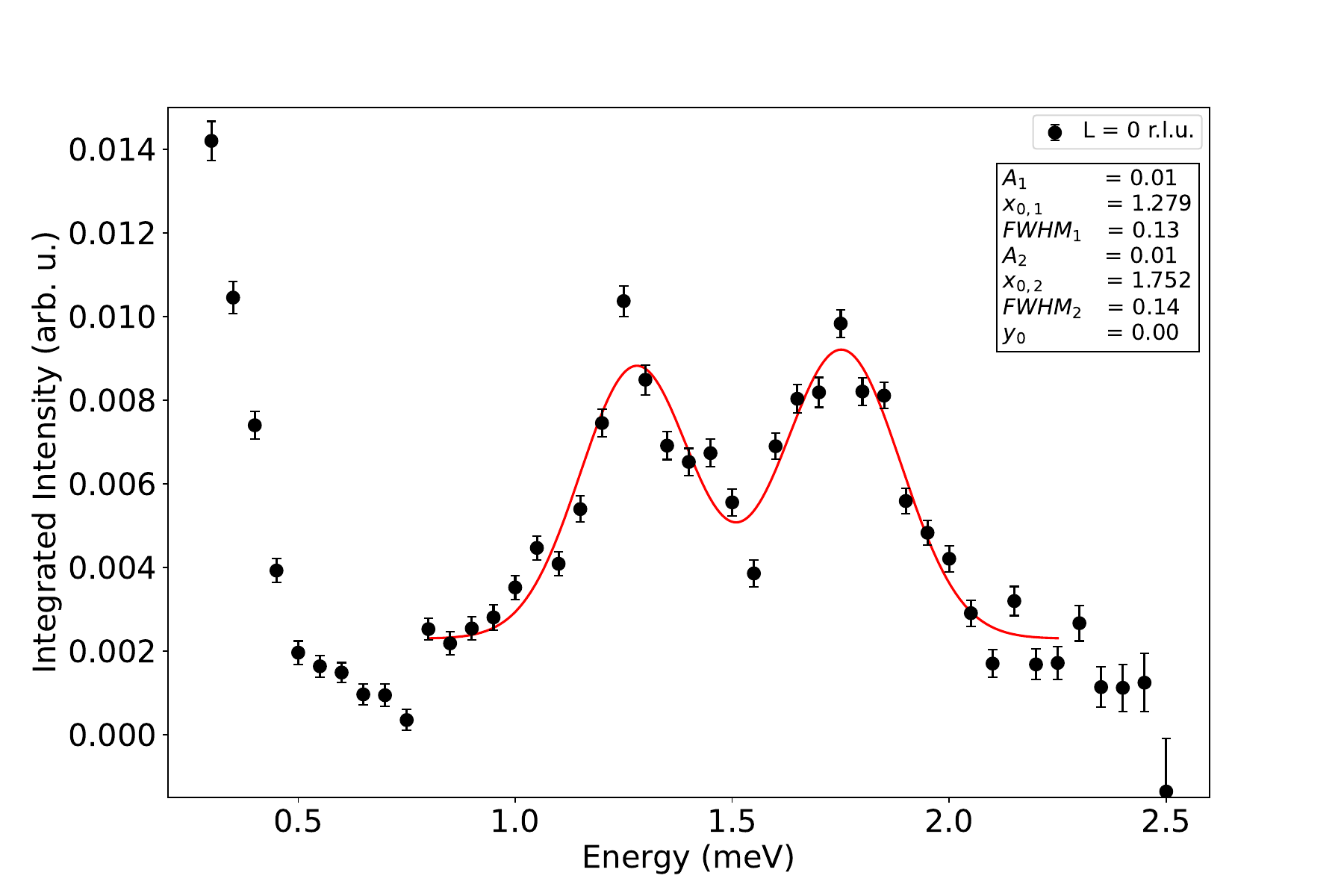}
		\caption{}
		\label{fig:L0}
	\end{subfigure}
	\begin{subfigure}[b]{0.32\textwidth}
		~
		\centering
		\includegraphics[width=1\textwidth,trim={0cm 0cm 0cm 0cm},clip]{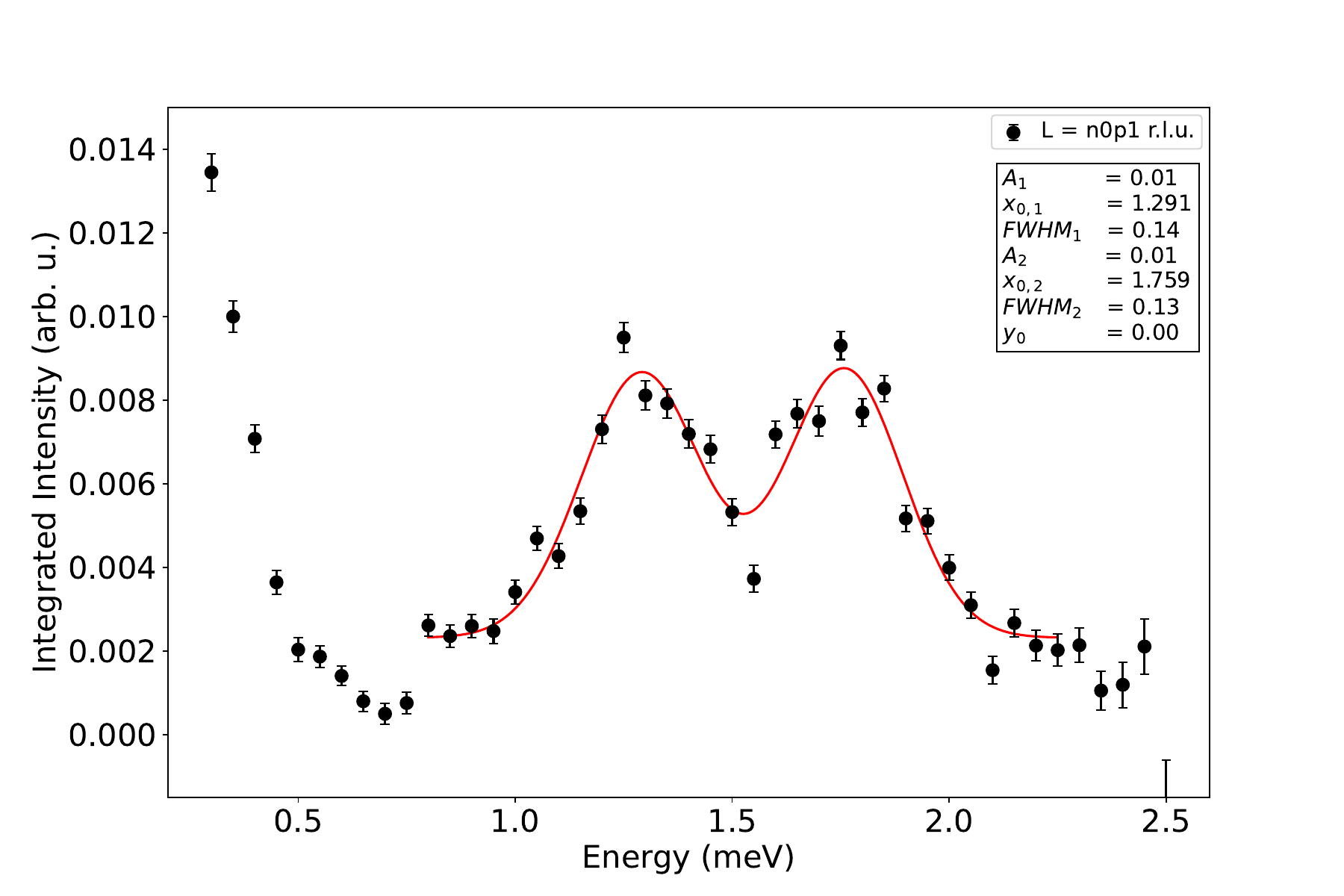}
		\caption{}
		\label{fig:Ln0p1}
	\end{subfigure}
	\begin{subfigure}[b]{0.32\textwidth}
		~
		\centering
		\includegraphics[width=1\textwidth,trim={0cm 0cm 0cm 0cm},clip]{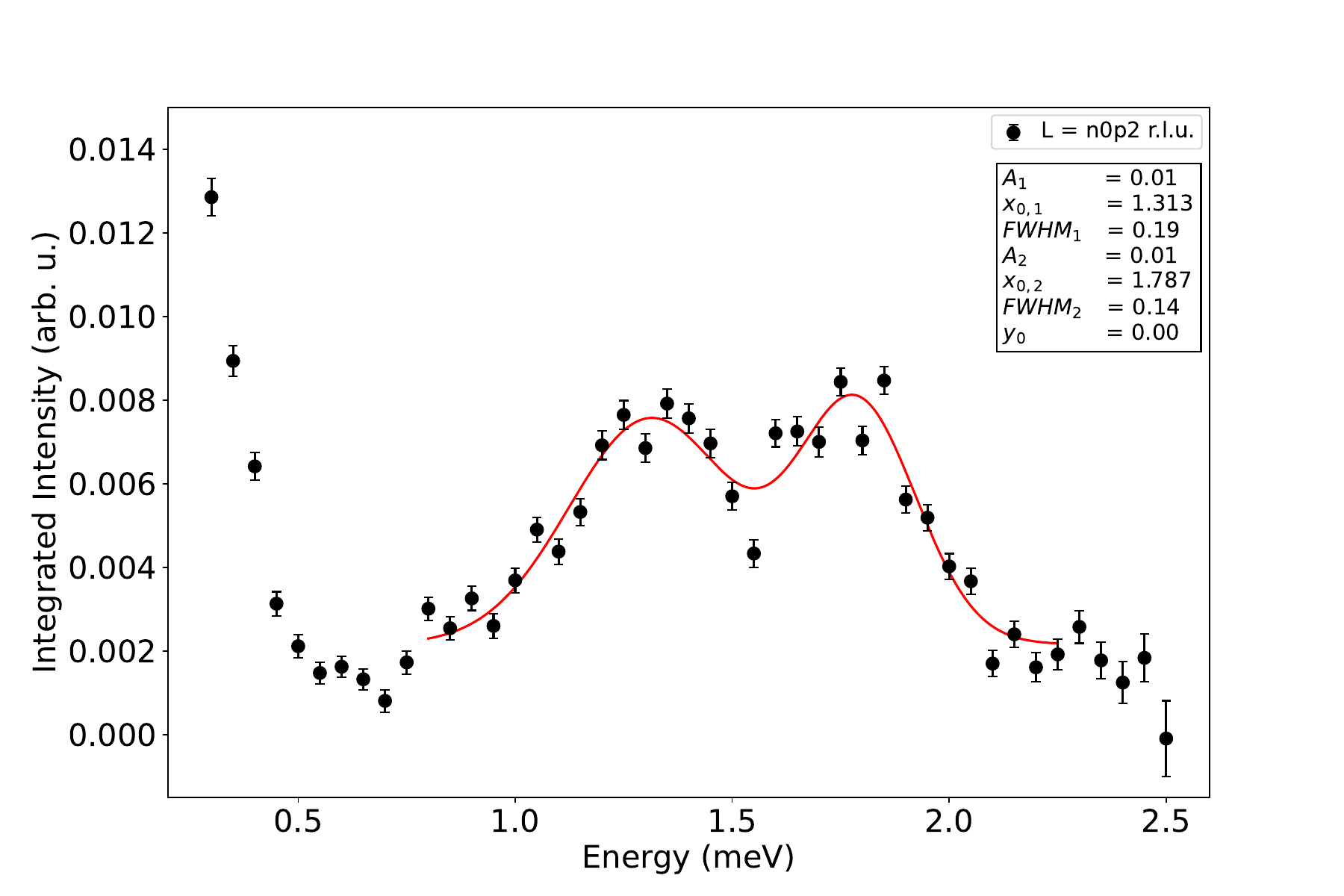}
		\caption{}
		\label{fig:Ln0p2}
	\end{subfigure}
	\begin{subfigure}[b]{0.32\textwidth}
	~
	\centering
	\includegraphics[width=1\textwidth,trim={0cm 0cm 0cm 0cm},clip]{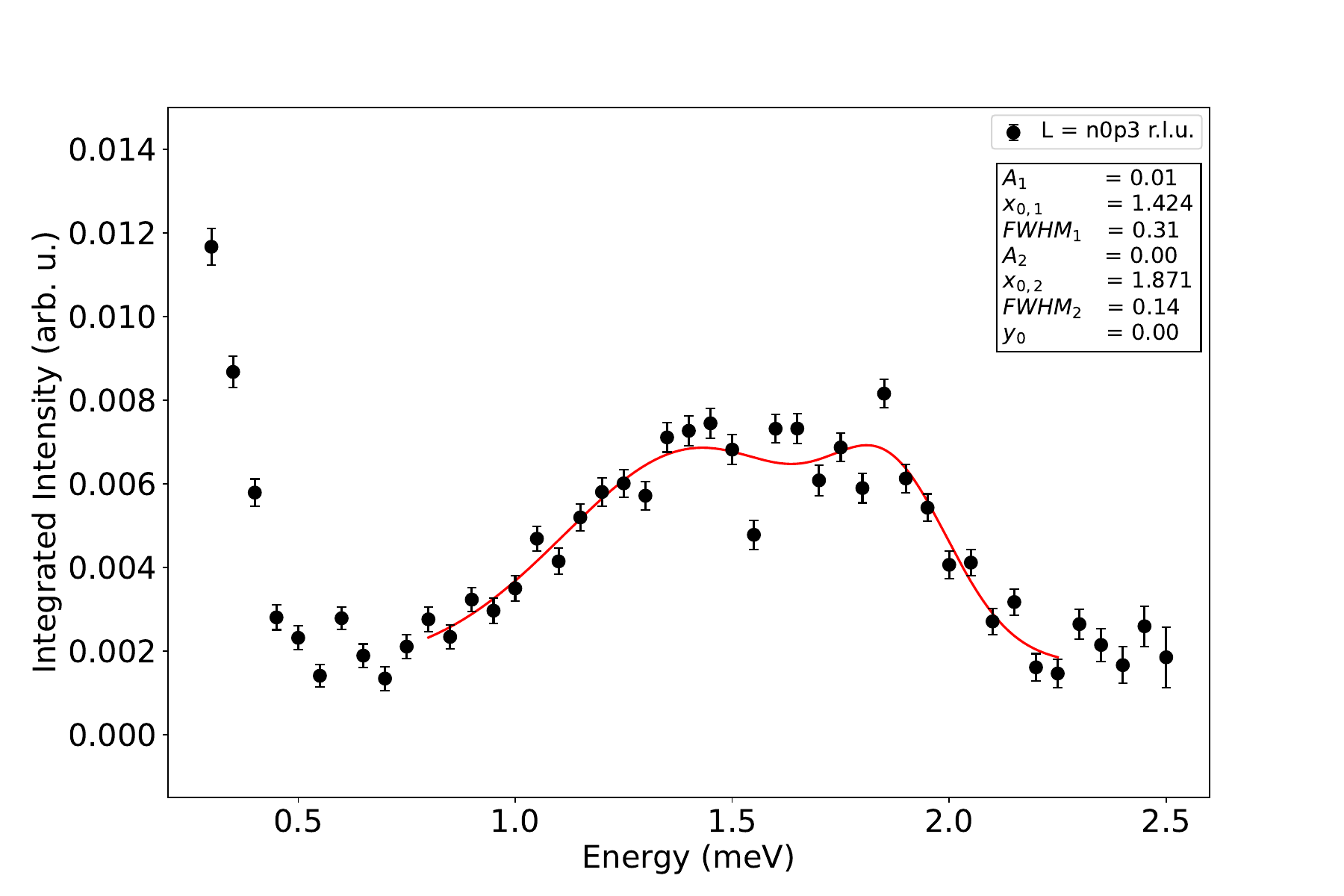}
	\caption{}
	\label{fig:Ln0p3}
	\end{subfigure}
	\begin{subfigure}[b]{0.32\textwidth}
	~
	\centering
	\includegraphics[width=1\textwidth,trim={0cm 0cm 0cm 0cm},clip]{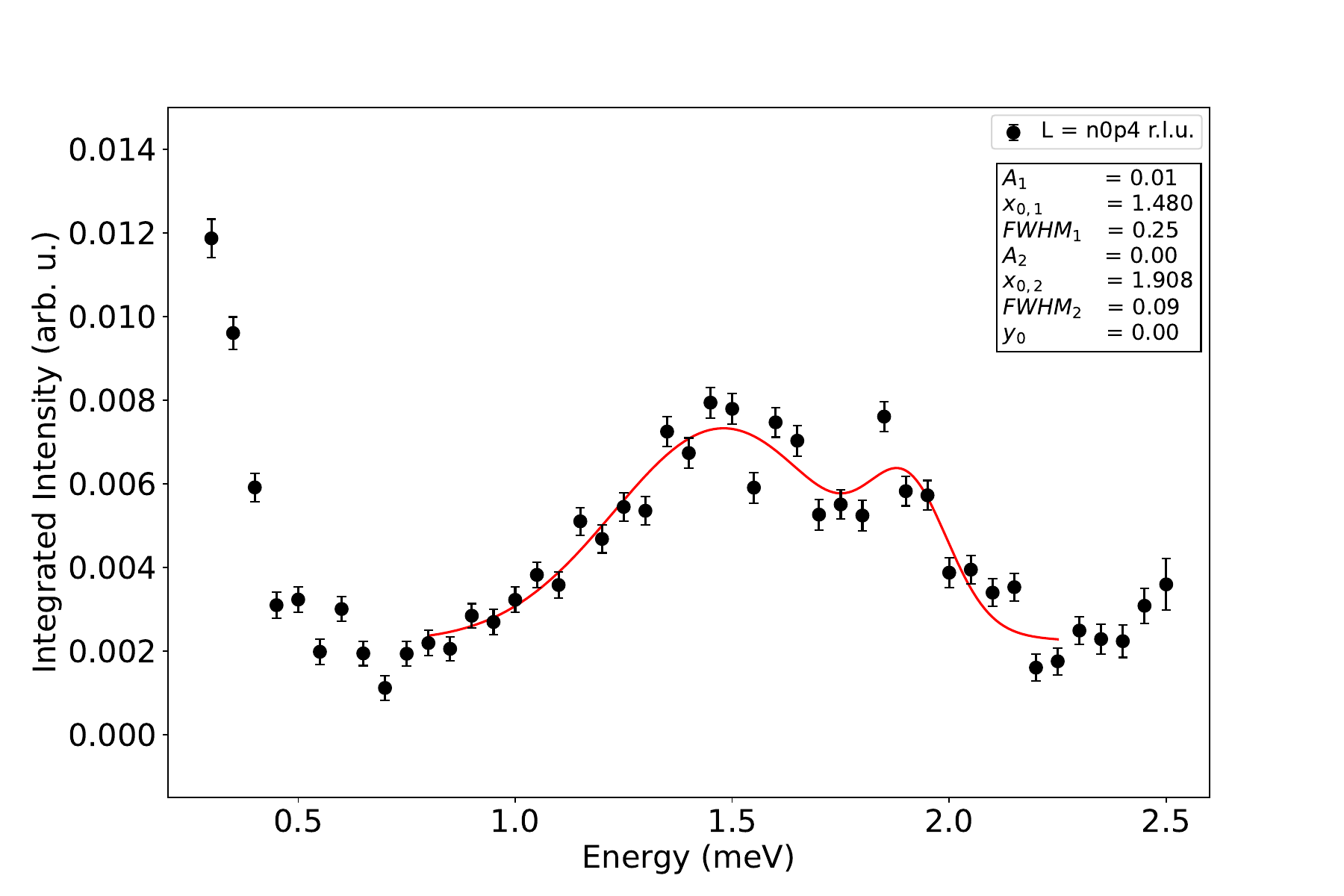}
	\caption{}
	\label{fig:Ln0p4}
	\end{subfigure}
	\begin{subfigure}[b]{0.32\textwidth}
	~
	\centering
	\includegraphics[width=1\textwidth,trim={0cm 0cm 0cm 0cm},clip]{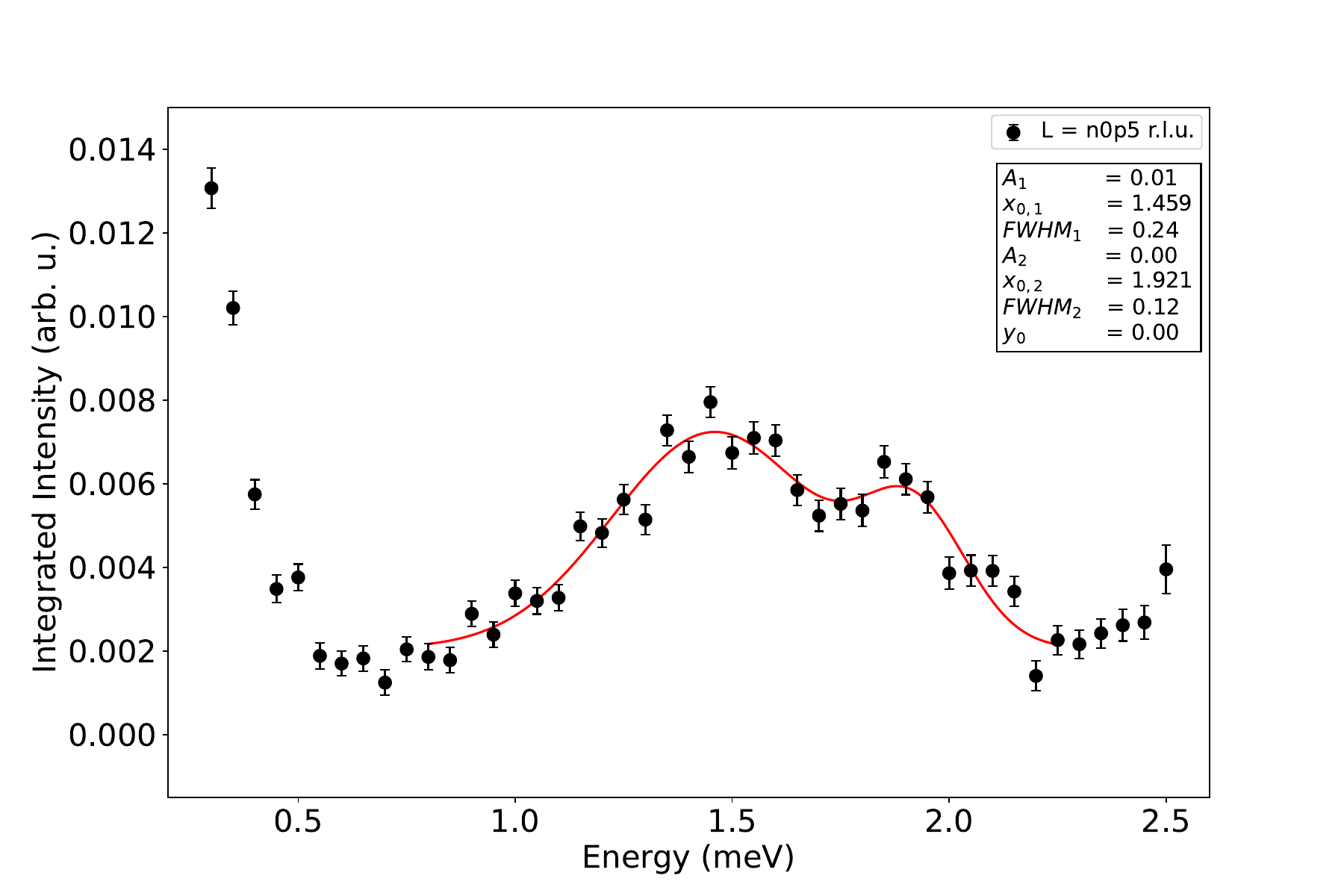}
	\caption{}
	\label{fig:Ln0p5}
	\end{subfigure}
	\begin{subfigure}[b]{0.32\textwidth}
	~
	\centering
	\includegraphics[width=1\textwidth,trim={0cm 0cm 0cm 0cm},clip]{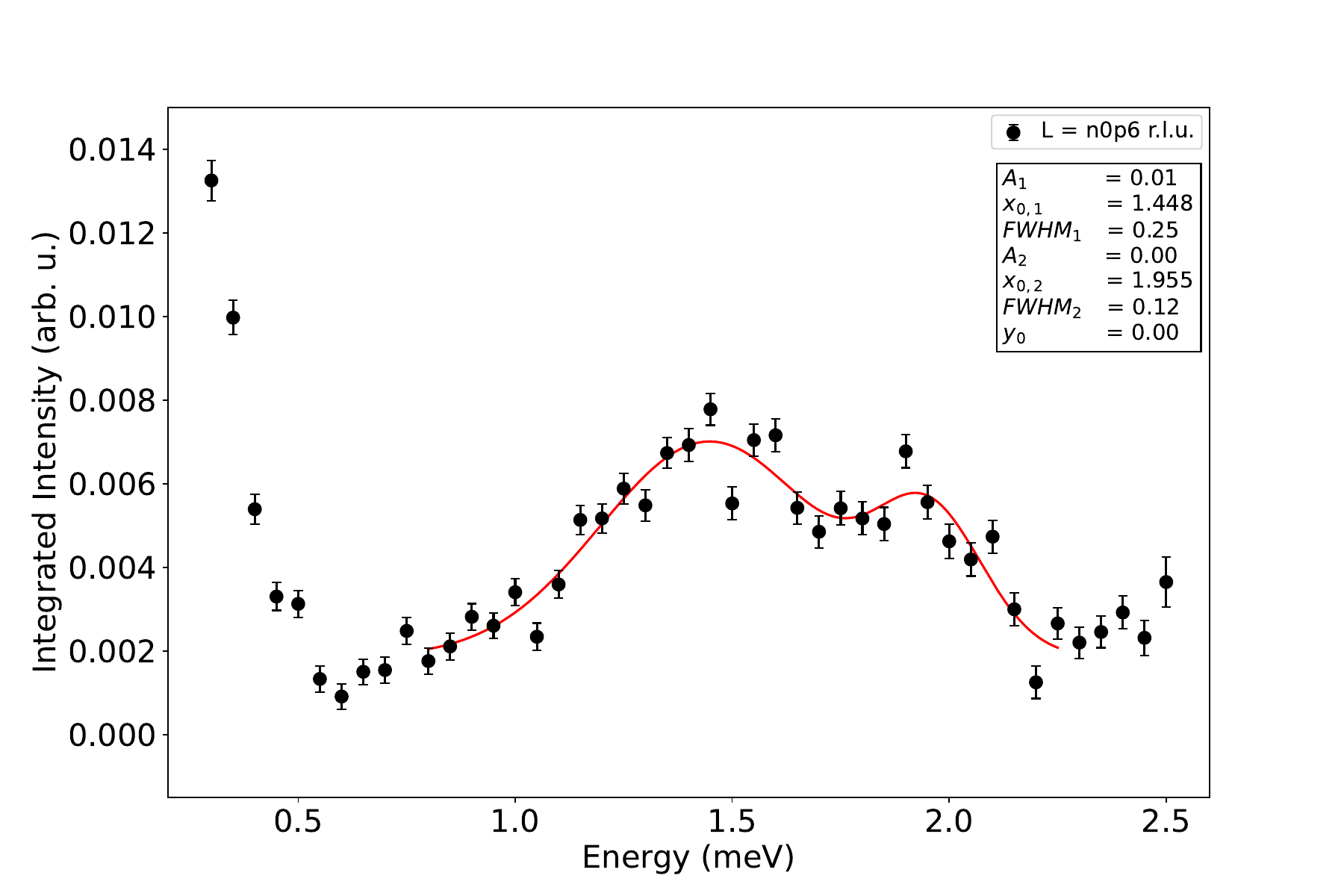}
	\caption{}
	\label{figLn0p6}
	\end{subfigure}
	\begin{subfigure}[b]{0.32\textwidth}
	~
	\centering
	\includegraphics[width=1\textwidth,trim={0cm 0cm 0cm 0cm},clip]{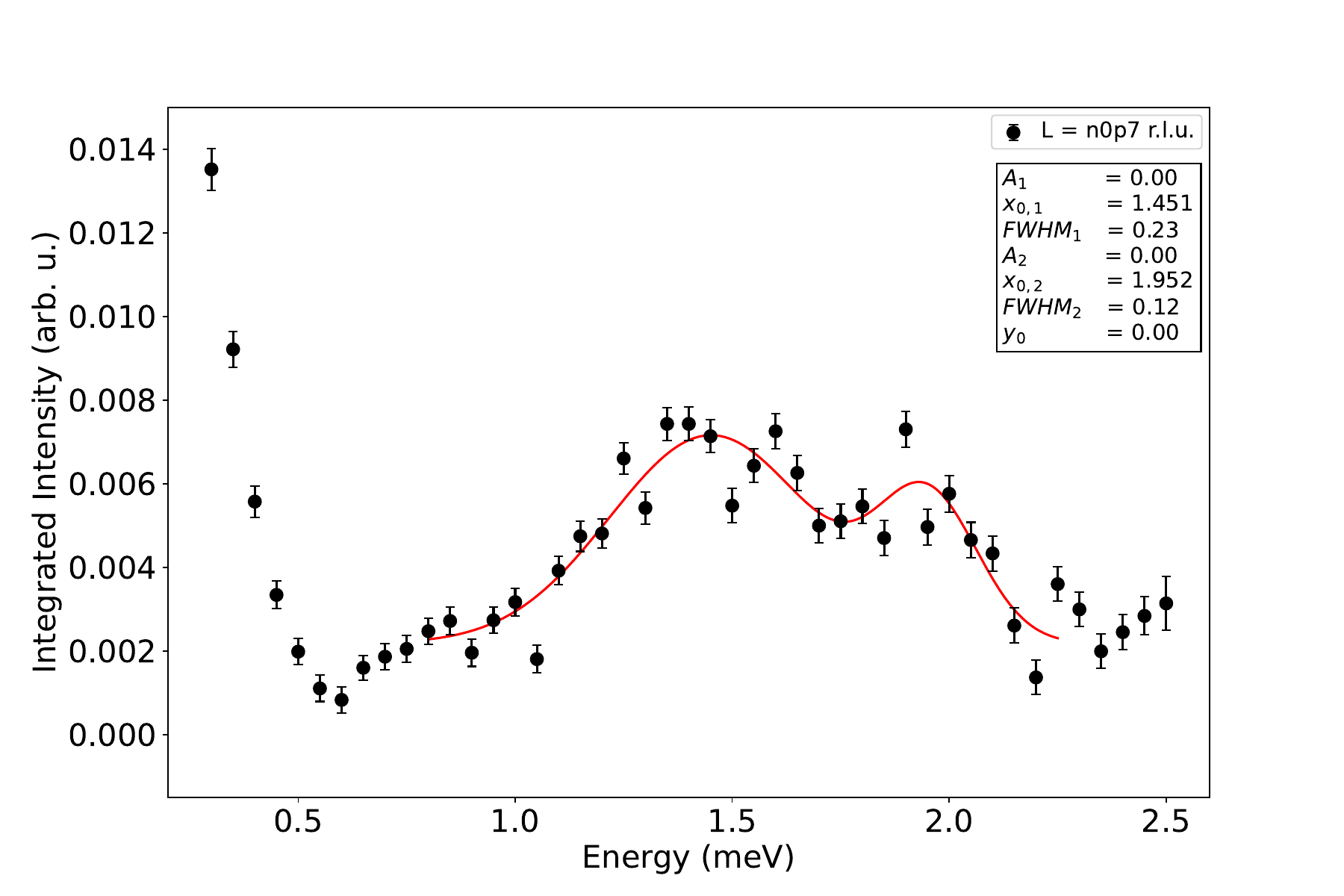}
	\caption{}
	\label{fig:Ln0p7}
	\end{subfigure}
	\begin{subfigure}[b]{0.32\textwidth}
	~
	\centering
	\includegraphics[width=1\textwidth,trim={0cm 0cm 0cm 0cm},clip]{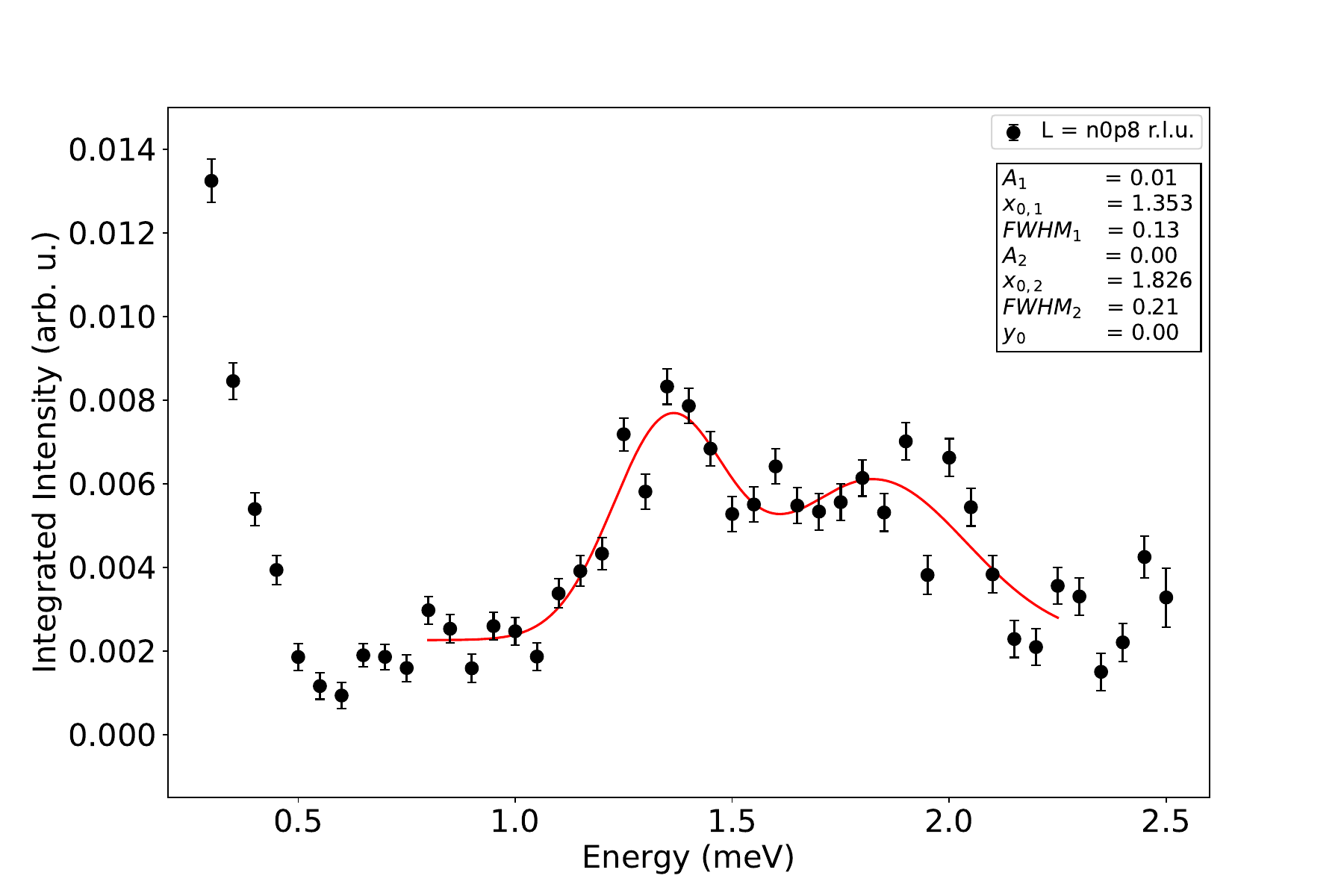}
	\caption{}
	\label{fig:Ln0p8}
	\end{subfigure}
	\begin{subfigure}[b]{0.32\textwidth}
	~
	\centering
	\includegraphics[width=1\textwidth,trim={0cm 0cm 0cm 0cm},clip]{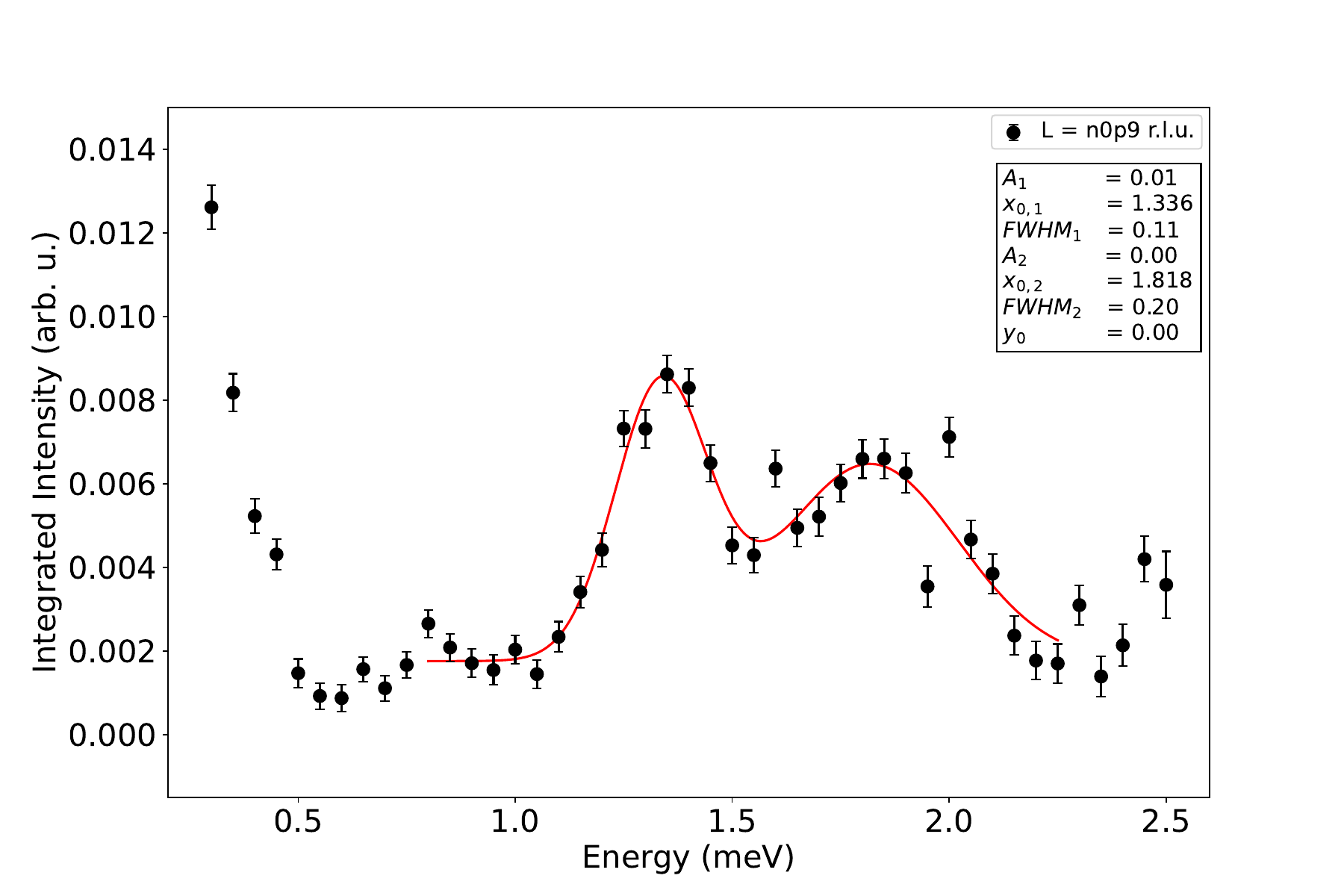}
	\caption{} 
	\label{fig:Ln0p9}
	\end{subfigure}
	\begin{subfigure}[b]{0.32\textwidth}
	~
	\centering
	\includegraphics[width=1\textwidth,trim={0cm 0cm 0cm 0cm},clip]{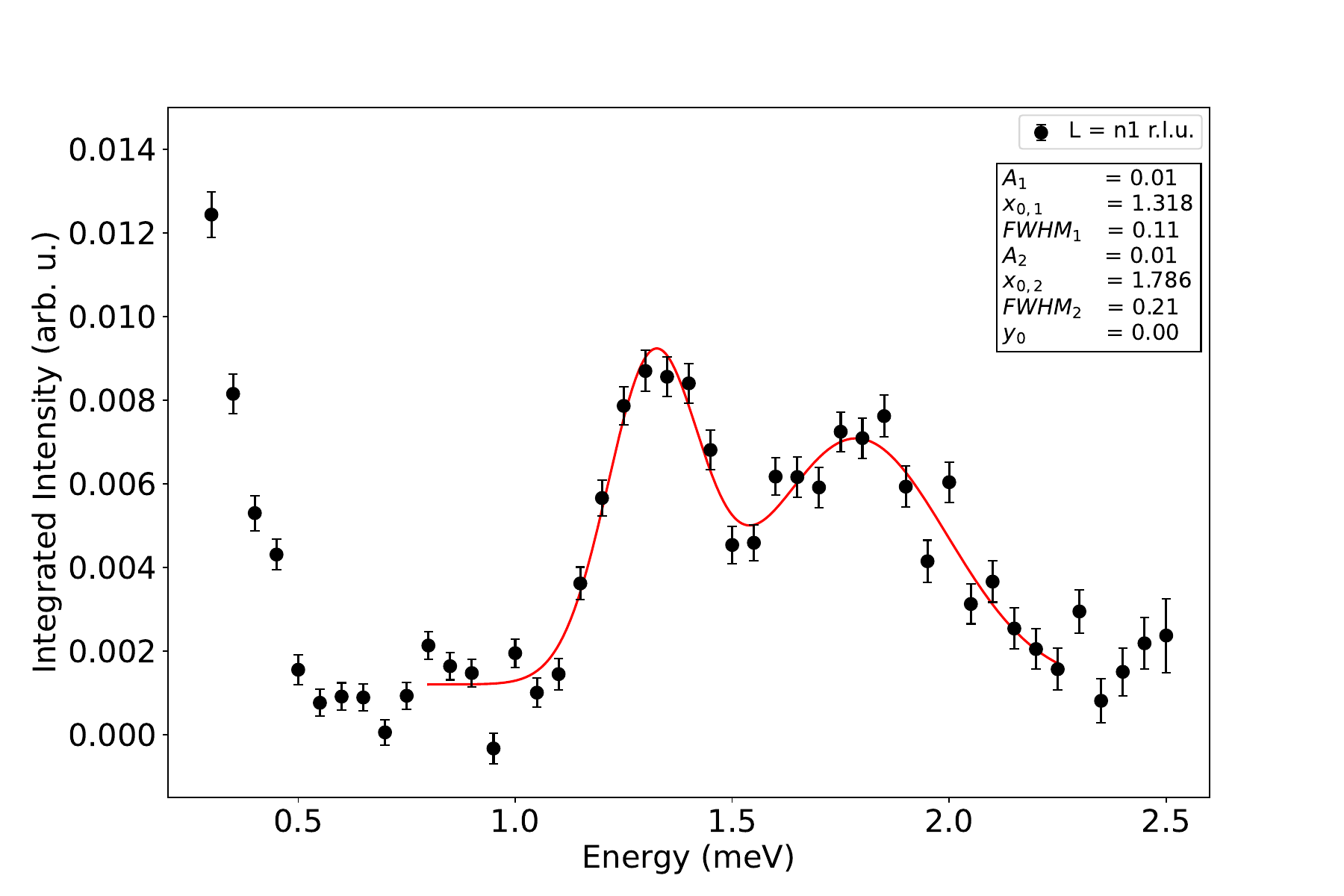}
	\caption{}
	\label{fig:Ln1}
	\end{subfigure}
	\caption{\subD{1D cuts}\addD{Constant-$Q$ cuts} of data at \addD{H=}0.5\,r.l.u. \subD{in the H direction}progressing with points in \subD{the L direction}\addD{L}. For each\subD{1D cut}, \addD{constant-$Q$ cut} the integration region in L is 0.2\,r.l.u., for H is 0.16\,r.l.u., and an energy interval of 0.05\,meV is used.}
	\label{fig:1D_cuts}
\end{figure}

\subD{1D cuts}\addD{Constant-$Q$ cuts} of the data at \addD{H=}0.5\,r.l.u. in \subD{the H direction} for various points in the L direction between 0\,r.l.u. and --1\,r.l.u. are shown in Fig. \ref{fig:1D_cuts}. The 1D cut shown in the main manuscript instead integrates over all of \subD{the L direction}\addD{L}. 

\begin{figure}[h!]
	\centering
	\includegraphics[width=0.9\textwidth,trim={0cm 0cm 0cm 0cm},clip]{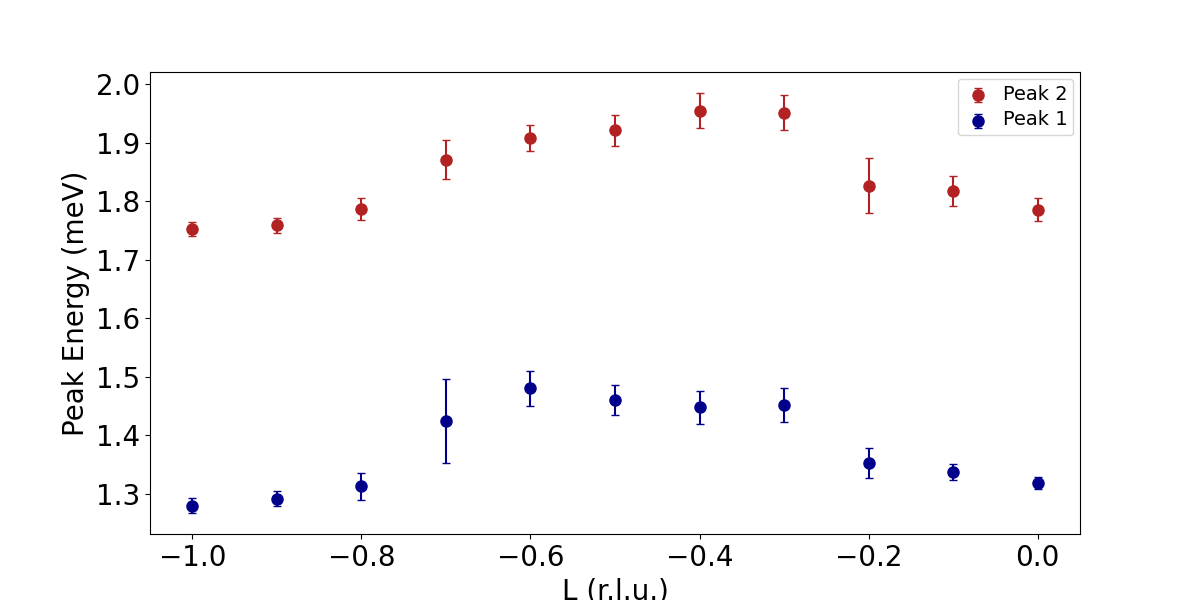}
	\caption{Plot of the peak positions in \subD{the H direction}\addD{H} at the magnetic zone center (H = 0.5\,r.l.u.) through slices in \subD{the L direction} \addD{L}.}
	\label{fig:peakpos}
\end{figure}

The \subD{1D cuts}\addD{constant-$Q$ cuts} through a progression in L maintain the same two distinctive spin-wave peaks at approximately 1.3\,meV and 1.75\,meV across the entire range in L between two magnetic zone boundaries. However, while the two spin-wave peaks persist across the full range in the L direction, they are not perfectly static. Fig. \ref{fig:peakpos} shows the progression of the peak positions in energy as a function of the point taken along L. The lower-energy mode progresses from 1.3meV at the magnetic zone boundary to 1.5\,meV at the magnetic zone center --- a shift of 0.2\,meV. The high-energy mode also shifts by 0.2\,meV from 1.8\,meV to 2.0meV from the magnetic zone boundary to the magnetic zone center. 

In order to confidently integrate along a particular reciprocal lattice direction, one must generally be confident that there is no dispersion in any of the excitations along that direction. However, the aforementioned main dataset has been processed with a full integration along \subD{the L direction}\addD{L}. There is a significant advantage to integrating over larger areas of reciprocal space in that one gains much stronger statistics for the dataset. While there is some dispersion along \subD{the L direction}\addD{L} for the modes observed along the \subD{the H direction}\addD{H}, the 0.2\,meV shift has been deemed to be relatively insignificant to overall character of the modes when viewed along \subD{the H direction}\addD{H} in comparison to the enhanced statistics it brings. 

\begin{figure}[h]
	\centering
	\begin{subfigure}[b]{0.32\textwidth}
		\centering
		\includegraphics[width=1\textwidth,trim={1cm 9cm 5.5cm 9cm},clip]{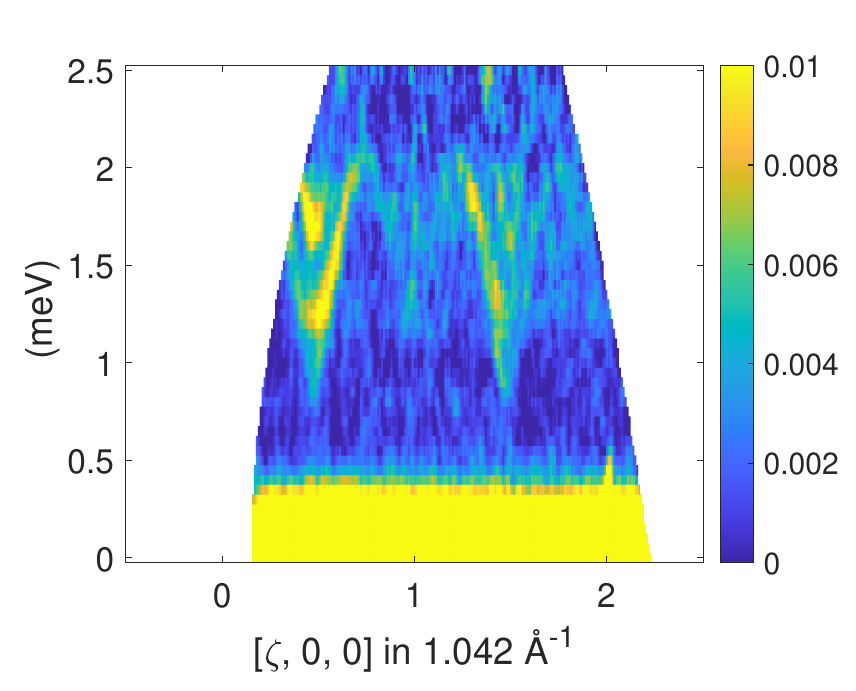}
		\caption{}
		\label{fig:2D_L0}
	\end{subfigure}
	\begin{subfigure}[b]{0.32\textwidth}
		~
		\centering
		\includegraphics[width=1\textwidth,trim={1cm 9cm 5.5cm 9cm},clip]{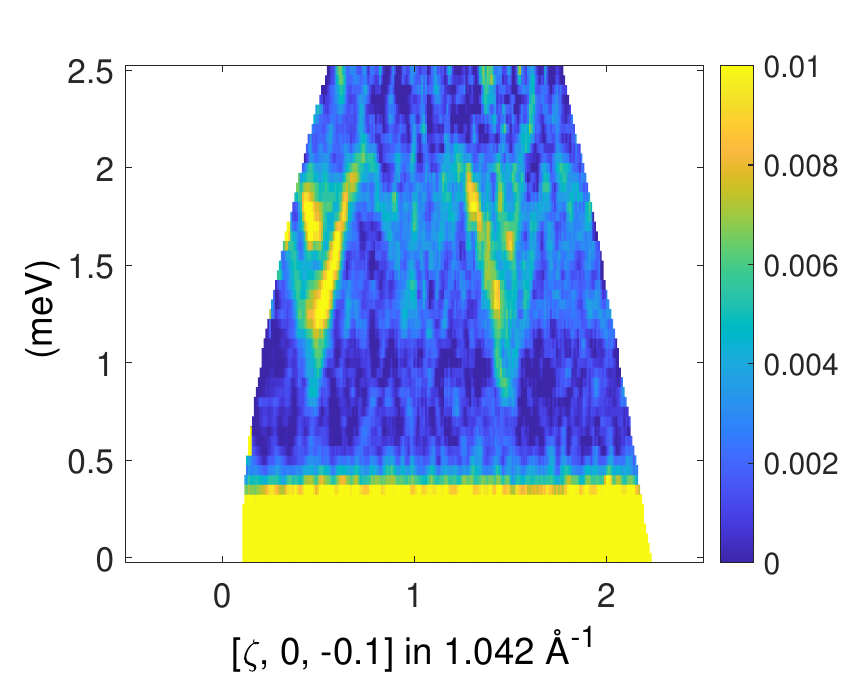}
		\caption{}
		\label{fig:2D_Ln0p1}
	\end{subfigure}
	\begin{subfigure}[b]{0.32\textwidth}
		~
		\centering
		\includegraphics[width=1\textwidth,trim={1cm 9cm 5.5cm 9cm},clip]{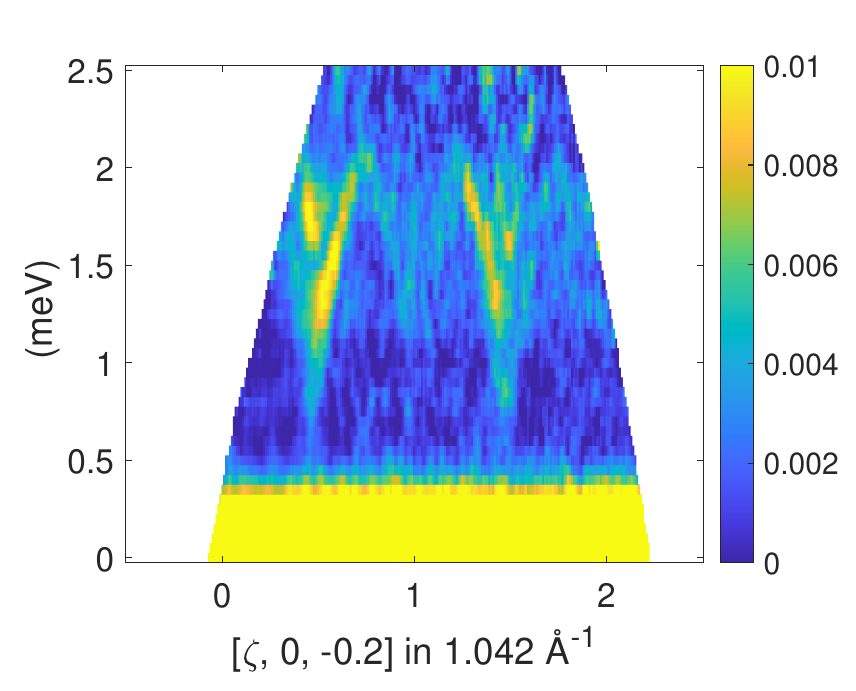}
		\caption{}
		\label{fig:2D_Ln0p2}
	\end{subfigure}
	\begin{subfigure}[b]{0.32\textwidth}
		~
		\centering
		\includegraphics[width=1\textwidth,trim={1cm 9cm 5.5cm 9cm},clip]{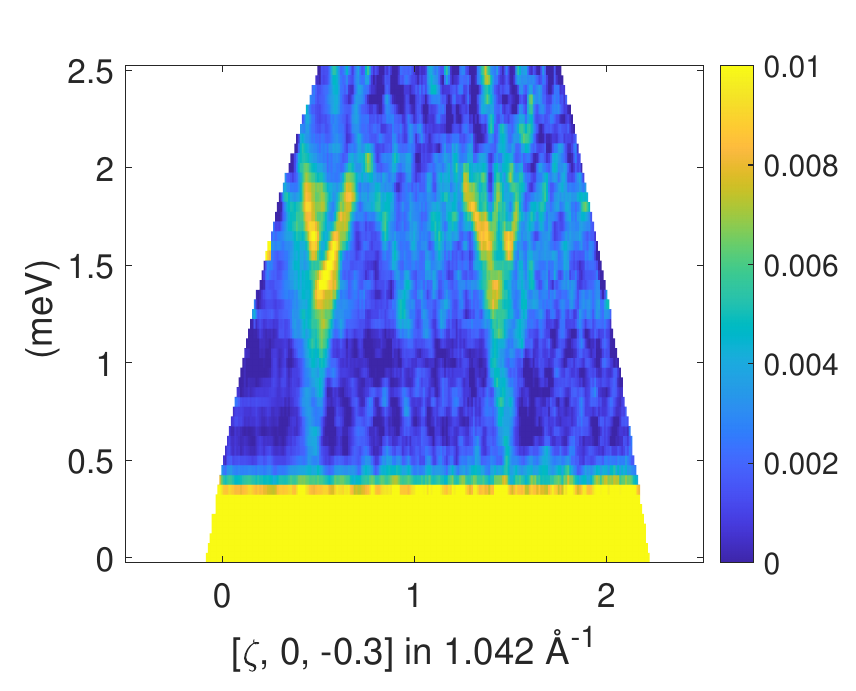}
		\caption{}
		\label{fig:2D_Ln0p3}
	\end{subfigure}
	\begin{subfigure}[b]{0.32\textwidth}
		~
		\centering
		\includegraphics[width=1\textwidth,trim={1cm 9cm 5.5cm 9cm},clip]{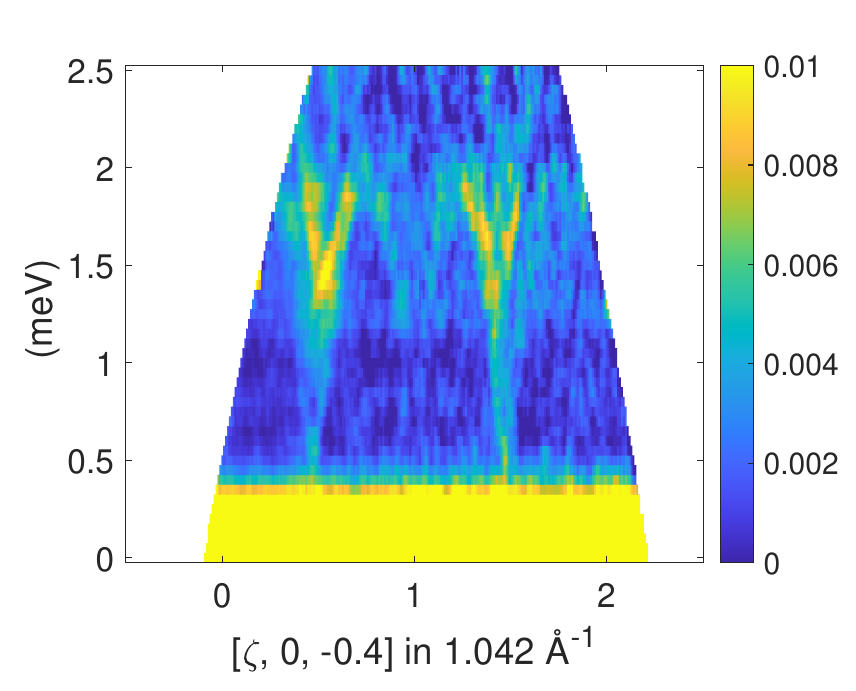}
		\caption{}
		\label{fig:2D_Ln0p4}
	\end{subfigure}
	\begin{subfigure}[b]{0.32\textwidth}
		~
		\centering
		\includegraphics[width=1\textwidth,trim={1cm 9cm 5.5cm 9cm},clip]{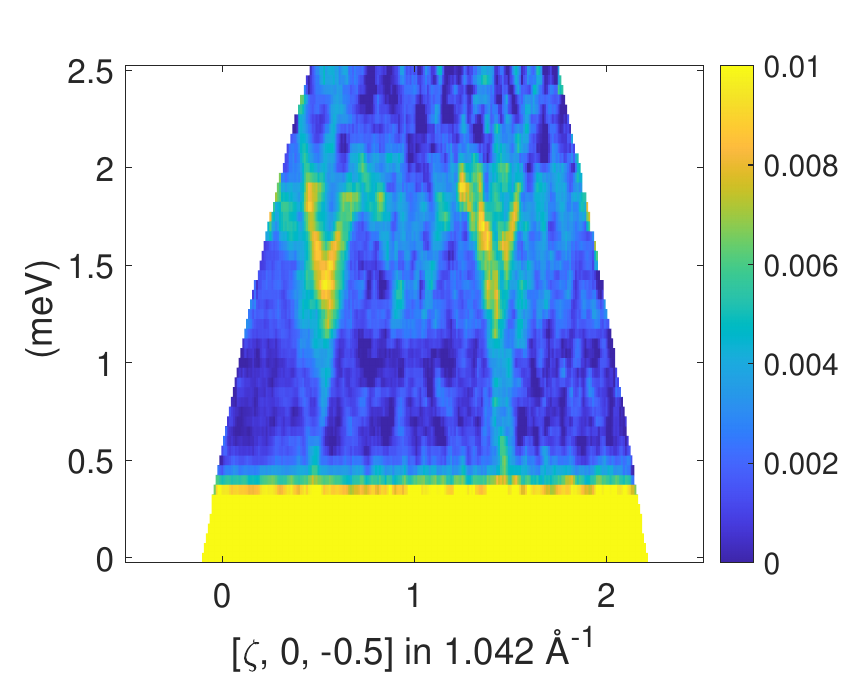}
		\caption{}
		\label{fig:2D_Ln0p5}
	\end{subfigure}
	\begin{subfigure}[b]{0.32\textwidth}
		~
		\centering
		\includegraphics[width=1\textwidth,trim={1cm 9cm 5.5cm 9cm},clip]{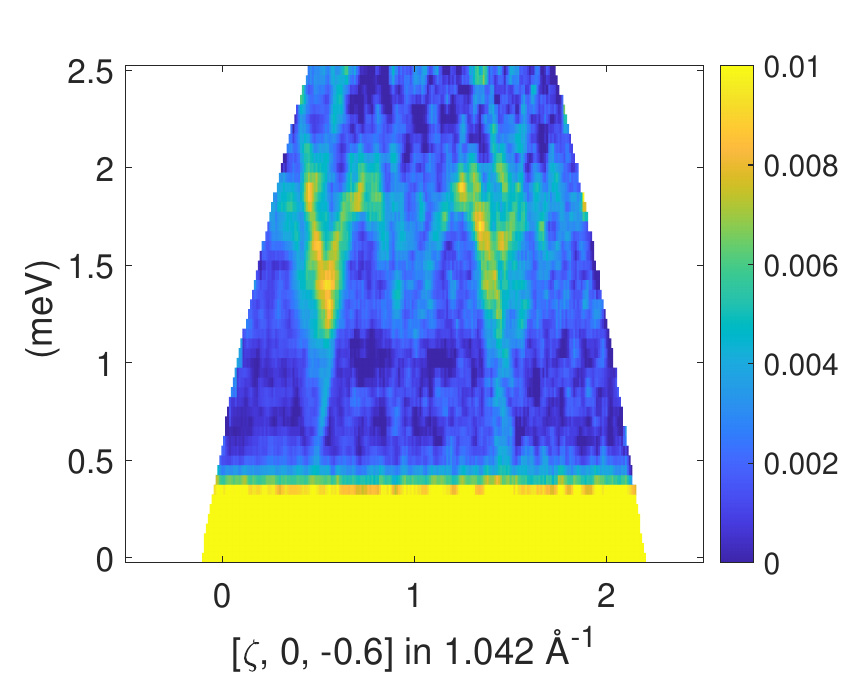}
		\caption{}
		\label{fig:2D_Ln0p6}
	\end{subfigure}
	\begin{subfigure}[b]{0.32\textwidth}
		~
		\centering
		\includegraphics[width=1\textwidth,trim={1cm 9cm 5.5cm 9cm},clip]{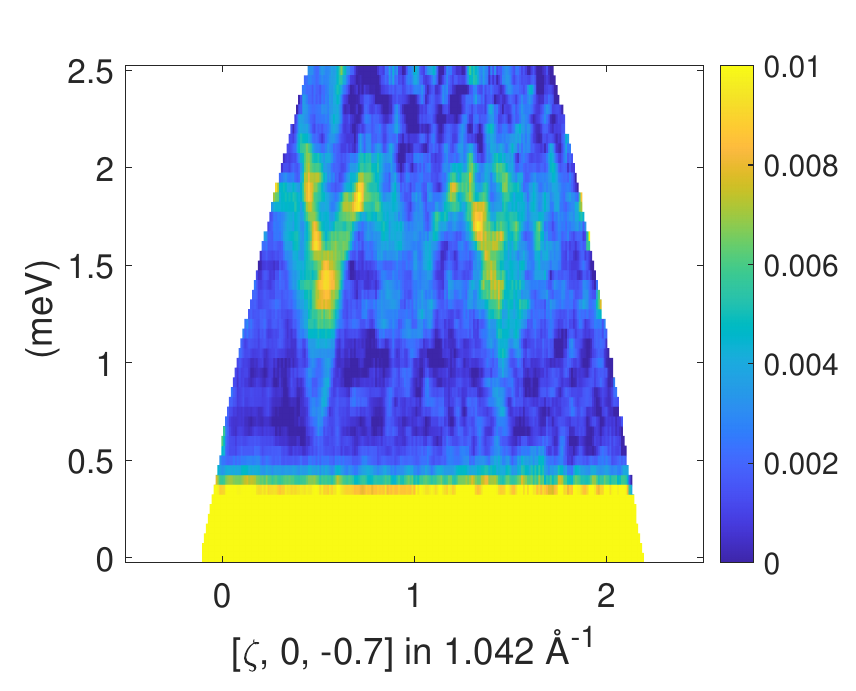}
		\caption{}
		\label{fig:2D_Ln0p7}
	\end{subfigure}
	\begin{subfigure}[b]{0.32\textwidth}
		~
		\centering
		\includegraphics[width=1\textwidth,trim={1cm 9cm 5.5cm 9cm},clip]{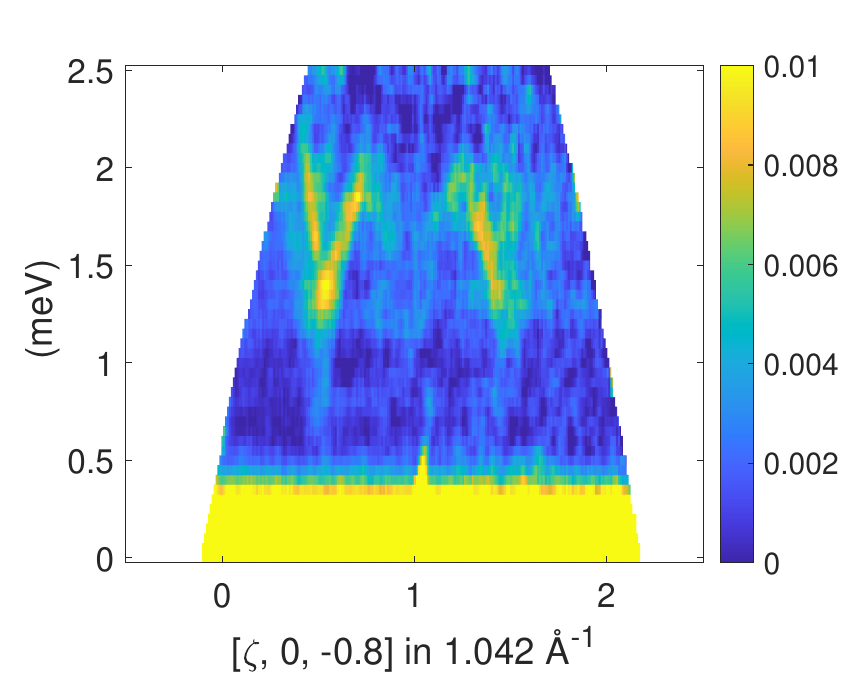}
		\caption{}
		\label{fig:2D_Ln0p8}
	\end{subfigure}
	\begin{subfigure}[b]{0.32\textwidth}
		~
		\centering
		\includegraphics[width=1\textwidth,trim={1cm 9cm 5.5cm 9cm},clip]{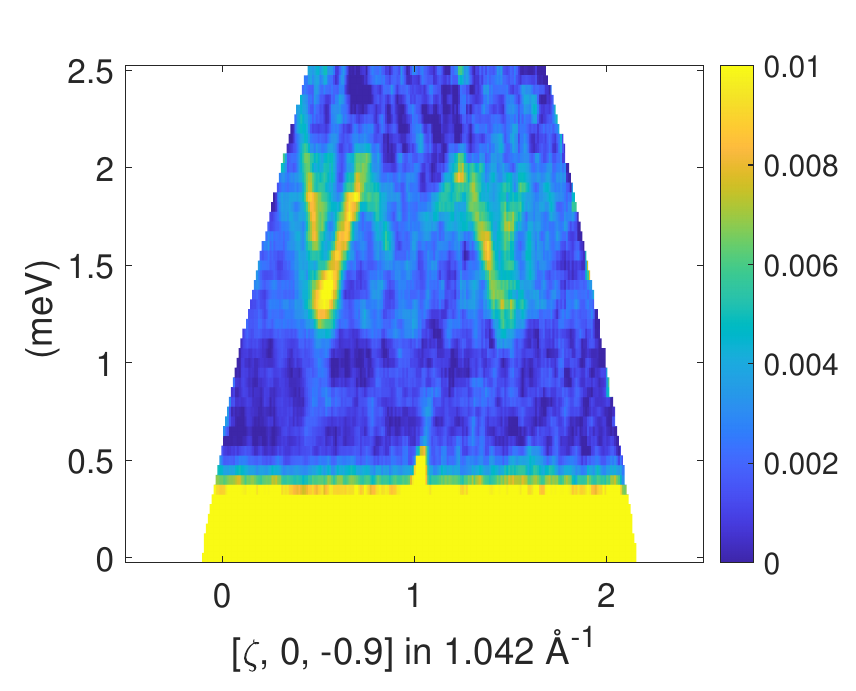}
		\caption{} 
		\label{fig:2D_Ln0p9}
	\end{subfigure}
	\begin{subfigure}[b]{0.32\textwidth}
		~
		\centering
		\includegraphics[width=1\textwidth,trim={1cm 9cm 5.5cm 9cm},clip]{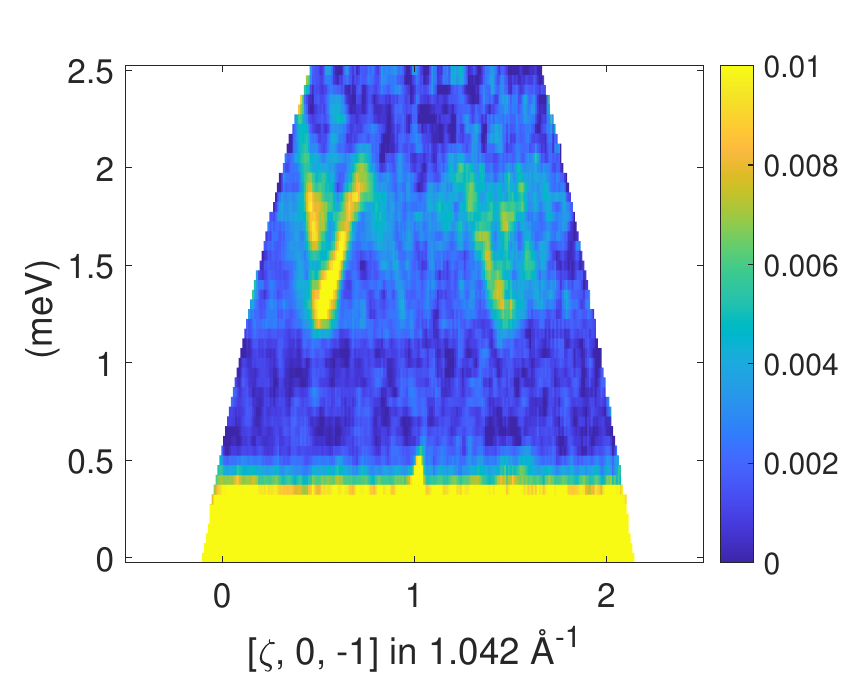}
		\caption{}
		\label{fig:2D_Ln01}
	\end{subfigure}
	\caption{2D cuts of data along H with each plot representing a progression in the L direction. For each 2D cut, the integration region in L is 0.2\,r.l.u., and an energy interval of 0.05\,meV is used.}
	\label{fig:2D_cuts}
\end{figure}

2D plots of the INS data showing spin-waves modes along the H direction for various points in the L direction between 0\,r.l.u. and --1\,r.l.u. are shown in Fig. \ref{fig:2D_cuts}. These correspond to the same points in L as shown in the 1D cuts.

The 2D plots can be used to gain confidence in the relatively static nature of the spin-wave modes seen along the H direction as a function of L. However, a subtlety not obvious in the fully L-integrated dataset nor the 1D cuts becomes apparent in the L-dependent 2D plots along the H direction. As L approaches --0.5\,r.l.u., a third `V'-shaped mode appears below the mode at 1.3\,meV. This weaker third mode moves down to the elastic line at 0.5\,meV at the magnetic zone center and does not appear to move any lower. This indicates that, despite the existence of a lower-energy mode than the two previously discussed, the system is still gapped. However, it is notable that a third mode not calculated in the LSWT appears in the INS data. The disagreement suggests that the larger complexity of the full exchange structure, beyond the zigzag-chain model, is needed to robustly describe the spin-wave modes along the H direction. 

\newpage
\clearpage

\section{Further Clarification on the Gap}

\begin{figure}[h!]
	\centering
	\includegraphics[width=0.99\textwidth,trim={0cm 0cm 0cm 0cm},clip]{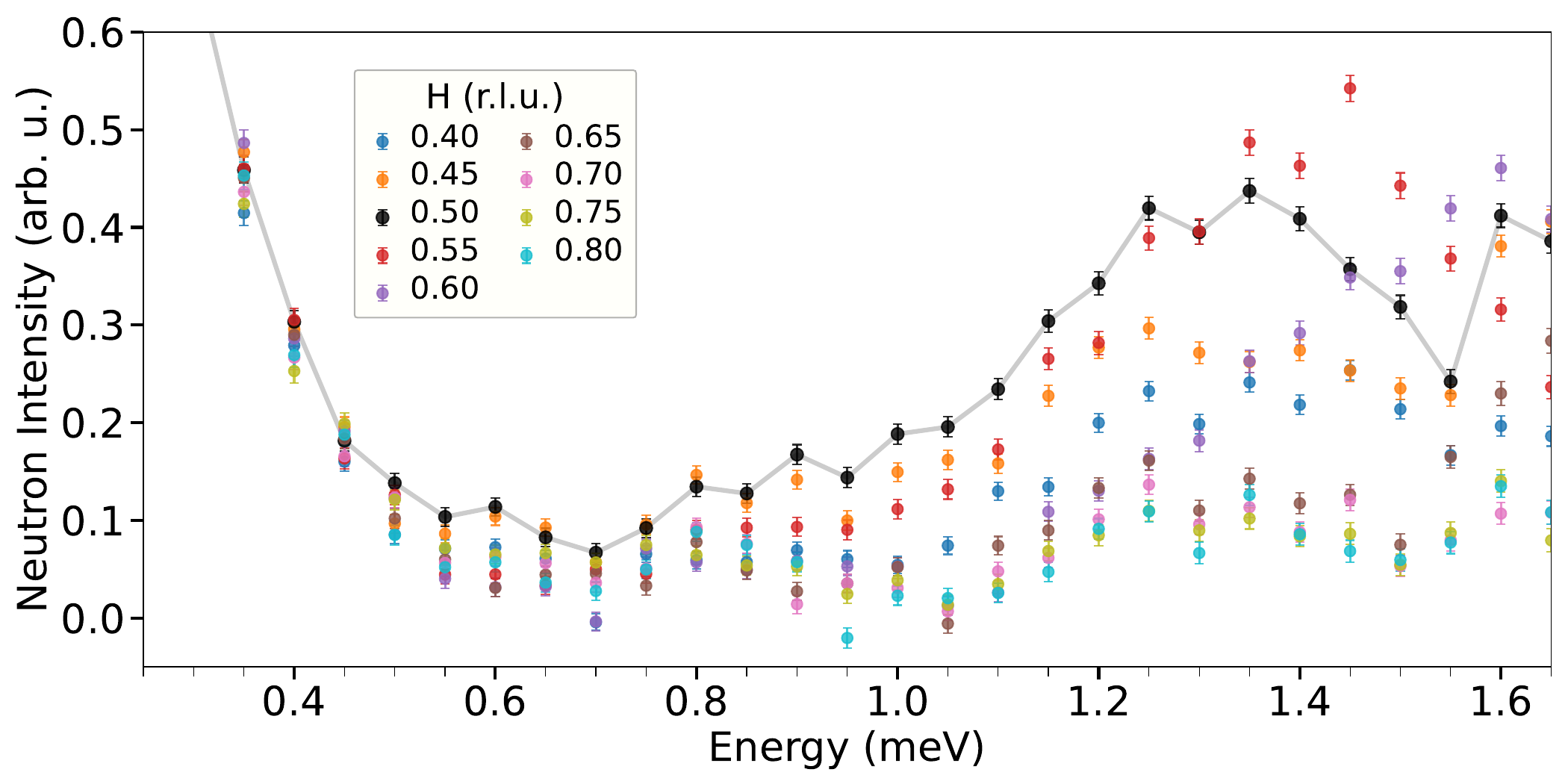}
	\caption{\addB{1D cuts along energy of the INS data at a range of H-points. All cuts are with an integration width of Q = 0.04\,r.lu. in H. The 1.3\,meV peak at H = 0.50\,r.l.u. progresses to higher energies as H increases or decreases from H = 0.50\,r.l.u. (the magnetic zone center). The cut at H = 0.50\,r.l.u. has been drawn thicker for contrast. Counts are within error of each other within the region below $\sim$0.75\,meV. This region therefore represents the background counts. Statistical variation is observed across the background counts.}
	\label{fig:gap_overlay}
}
\end{figure}

To confidently quantify the energy gap in atacamite along the H direction, a comparison against the background signal must be considered. \addB{This comaprison has been presented in Fig. 4 of the main text. This supplemental section provides further discussion about the process and the robustness of the gap determination.}

\addB{Fig. \ref{fig:gap_overlay} presents a set of 1D cuts along energy for a range of H-points. At H = 0.50\,r.l.u., the 1.3\,meV excitation peak is present and as H increases or decreases from this point, the excitation peak increases in energy, leaving a wider region of background counts. The counts are generally within error of each inside the region below $\sim$0.75\,meV. }

\addB{In the inset of Fig. 4 of the main text showing the subtraction between two 1D cuts at H = 0.50\,r.l.u. and H = 0.75\,r.l.u., the lowest-energy point which is within error of zero counts in the subtraction is also observed at $0.75\,$meV. However, the peak of the associated excitation is centered at $1.3\,$meV. The fact that the difference from the background counts does not reach zero until $0.75\,$meV is likely contributed to by peak broadening from lifetime effects, imperfect sample co-alignment, and some non-concentricity of the sample with respect to the neutron beam. Thus, the true gap is somewhere between $0.75\,$meV and $1.3\,$meV, but, given the experimental limitations, it can only be confidently quantified to be at least $0.75\,$meV, as stated in the main text.}


\sub{Fig. S10 (a) and Fig. S10 (b) consider two 1D cuts. One is along H = 0.50\,r.l.u. at the magnetic zone center. The other is along H = 0.75\,r.l.u., a region where the main modes have dispersed to higher energies leaving the low energy region (below $\sim 1.5\,$meV) equivalent to the background.}

\sub{Fig. S10 (b) presents a subtraction between the 1D excitation spectrum at the magnetic zone center to the spectrum at  H = 0.75\,r.l.u. in order to measure the gap. The lowest-energy point which is within error of zero counts in the subtraction is at $0.75\,$meV. However, the peak of the associated excitation is centered at $1.3\,$meV. The fact that the difference from the background counts does not reach zero until $0.75\,$meV is likely contributed to by peak broadening from lifetime effects, imperfect sample co-alignment, and some non-concentricity of the sample with respect to the neutron beam. Thus, the true gap is somewhere between $0.75\,$meV and $1.3\,$meV, but, given the experimental limitations, it can only be confidently quantified to be at least $0.75\,$meV.}

\newpage
\clearpage

\section{Validity of LSWT on the $J_1-J_2$ Chain}
A reasonable concern to have is that modeling a quantum mechanical magnetic system such as the spin-1/2 zigzag chain with LSWT, which is only a first-order approximation for the solution of a Hamiltonian, will be problematic. A specific concern would be that if the system exhibited a gap which is quantum in nature, then LSWT would not capture this. Additionally, the dispersion behavior may be entirely mischaracterized.

\begin{figure}[h]
	\centering
	\includegraphics[width=0.99\textwidth,trim={0cm 0cm 0cm 0cm},clip]{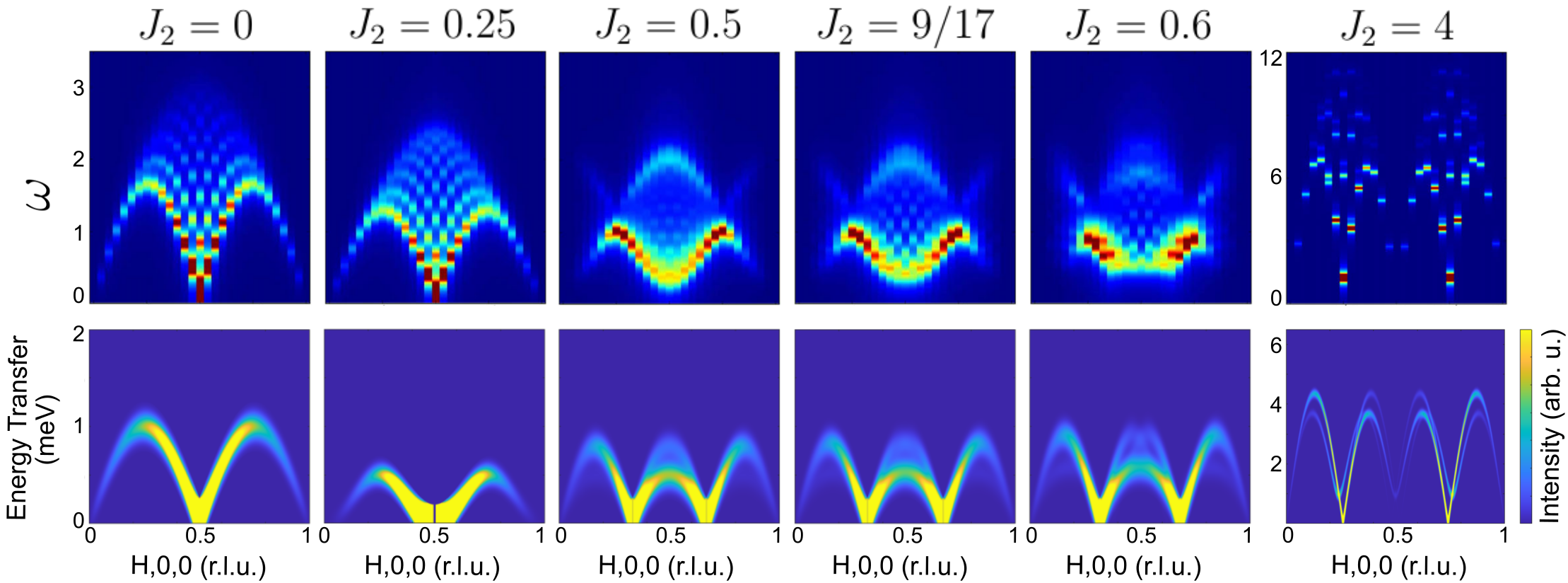}
	\caption{\addD{The top row} gives the ED calculations of \subD{the} the $J_1-J_2$ chain. These has been adapted from Ref. \cite{lavarelo_2014}. \addD{The bottom row} shows the LSWT calculations on the equivalent $J_1-J_2$ chain along the $a$-crystallographic direction, which follow the order of increasing $J_2$ as labeled in \addD{the top row}.}
	\label{fig:EDvsLSWT}
\end{figure}

Exact diagonalization (ED) offers a method for calculating the dynamics of spin excitation while maintaining higher-order accuracy in the diagonalization of the Hamiltonian, capturing the quantum effects. Previous work by A. Lavarélo and G. Roux \cite{lavarelo_2014} presents ED calculations of the spin-wave and spinon spectra for a general  $J_1-J_2$ chain for various $J_1/J_2$ ratios. Fig. \ref{fig:EDvsLSWT} shows the comparison between these ED calculations and equivalent LSWT calculations.

LWST is able to reproduce the dispersion characteristics, the intensity distribution, and the periodicity of the ED calculations for the $J_2 = 0, J_2 = 0.25$, and $J_2 = 4$ cases. However, the spinon continuum for $J_2 = 0$ and $J_2 = 0.25$ do not appear in LSWT, and the second mode in the LSWT calculation for $J_2 = 4$ does not appear in the ED calculation. More crucially, the spin-excitation spectra for $J_2 = 0.5, J_2 = 9/17$, and $J_2 = 0.6$ do not match at all between the two techniques. This is the region where the largest gap is expected \cite{lavarelo_2014, Steven_1996}, and the classical approach of LSWT fails to capture this gap. However, the zigzag-chain model proposed for atacamite in this work has an equivalent $J_2$ value of 1.9. 

Given that the gap is expected to decay exponentially with $J_2$, the energy gap at $J_2 = 1.9$ should be similar to the $J_2 = 4$ case. So, the zigzag-chain model is within a region for a $J_1-J_2$ chain where LSWT calculations arrive at similar results to ED, and LSWT represents a sensible approach for understanding the spin-wave spectra of atacamite when considering the zigzag-chain model.

%

\bibliography{ata_sup_2023}